\documentstyle[aps,preprint,pre]{revtex}

\begin{document}
\draft

\baselineskip 0.4cm

%\twocolumn[\hsize\textwidth%
%\columnwidth\hsize\csname@twocolumnfalse\endcsname

\title{\bf Many-body Effects on Excitonic Optical Properties of Photoexcited 
Semiconductor Quantum Wire Structures}

\author{D. W. Wang and S. Das Sarma} 

\address{Department of Physics, University of Maryland, College Park, Maryland 20742-4111}

\date{\today}
\maketitle
\pagenumbering{arabic}
\vspace*{1.5cm}

\begin{abstract} 
\baselineskip 0.4cm

We study carrier interaction induced 
many-body effects on the excitonic optical properties 
of highly photoexcited
one-dimensional semiconductor quantum wire systems by solving the dynamically
screened Bethe-Salpeter equation using realistic Coulomb interaction
between carriers.
Including dynamical screening effects in the electron/hole self-energy 
and in the electron-hole interaction vertex function, 
we find that the excitonic absorption is essentially peaked 
at a constant energy for a large range of photoexcitation density
($n= 0-6\times 10^5$ cm$^{-1}$), above which the absorption peak 
disappears without appreciable gain i.e., \textit{no} exciton to free 
electron-hole plasma Mott transition is observed,
in contrast to previous theoretical results but in agreement 
with recent experimental findings. 
This absence of gain (or the non-existence of a Mott transition)
arises from the strong inelastic scattering by one-dimensional 
plasmons or charge density excitations, closely related to 
the non-Fermi liquid nature of one-dimensional systems.
Our theoretical work demonstrates a transition or a crossover in 
one-dimensional photoexcited electron-hole system from an effective Fermi
liquid behavior associated with a dilute gas of noninteracting excitons
in the low density region ($n<10^5$ cm$^{-1}$)
to a non-Fermi liquid in the high density region ($n>10^5$ cm$^{-1}$).
The conventional quasi-static approximation for this problem
is also carried out to compare with the full dynamical results. 
Numerical results for exciton binding energy and absorption spectra 
are given as functions of carrier density and temperature.
\end{abstract}

\vfill\eject
\vskip 1pc
%\vskip 1pc]
%\narrowtext
%%%%%%%%%%%%%%%%%%%%%%%%%%%%%%%%%%%%%%%%%%%%%%%%%%%%%%%%%%%%%%%%%%%%%%%%%
\section{Introduction}
Excitons in low dimensional semiconductor systems have been extensively 
studied in the recent past. Present interest has focused on one-dimensional
excitons in artificially structured semiconductor quantum wire (QWR)
systems where spectacular improvements in growth and nanofabrication 
techniques have led to very narrow
wires of nanostructure size ($< 100$ \AA\ in GaAs) with rather
deep conduction band electron confinement energy
($\sim 150$ meV) and large conduction subband spacing ($\sim 20$ meV) 
\cite{exciton_exp_amb,exciton_exp_weg,exciton_laser,bgr_exp} 
so that the electrons in the conduction band
of such a QWR most likely form a pure one-dimensional (1D) system. 
For the holes in the QWR
valence band, the bare confinement potential (for example,
in GaAs-AlGaAs system) is known to be too shallow ($\sim 10$ meV) 
for a hole to be one-dimensionally
confined in these QWR structures. Including the
Coulomb interaction between electrons and holes along the transverse 
(i.e. perpendicular to the 1D free motion direction)
directions of the wire, however, Glutsch \textit{et al}.~\cite{glutsch}
find that even the holes in the valence band of QWR can be strongly 
localized in the transverse plane, leading to 
both electrons and holes being effectively 1D (or rather quasi-1D) in the
dynamical sense.
Therefore an exciton in such ultranarrow QWR nanostructures can 
be effectively thought of as
a bound pair of a 1D electron and a 1D hole 
with the carrier dynamics being free along the 1D wire direction
as long as one is interested in low energy (lower than confinement energy
$\sim 20-100$ meV) excitonic optical properties.
Such strong confinement for both electrons and holes also
substantially enhances the excitonic binding energy 
leading to novel optical phenomena.
In the low (electron-hole) density limit 
without considering the self-energy correction to
the conduction and the valence band energies as well as neglecting all
dynamical screening effects, the single electron and
single hole problem in forming the exciton can be exactly solved as
a quasi-1D hydrogenic (Wannier exciton) atom with an exciton radius of
about 100 \AA\ for GaAs based QWR systems.
This single 1D exciton problem, where an electron and a hole in a QWR 
form a bound excitonic state, has been studied extensively in the recent
literature in the context of understanding QWR excitonic optical properties.
Such a non-interacting exciton picture, based on a simple single particle
electron-hole hydrogenic bound state scenario, obviously only applies in 
the dilute low carrier density limit when the excitons or the bound 
electron-hole pairs are effectively very far from each other forming 
a noninteracting exciton gas. We will refer to this situation as a 
Fermi liquid (because in 1D only an effectively noninteracting system 
may behave as a Fermi liquid) or a noninteracting exciton gas. In 
the high carrier density situation the excitons must overlap a great 
deal, and the simple Fermi liquid picture of a noninteracting exciton 
gas would not apply. Our main goal in this paper is to theoretically 
study this transition between the low density (Fermi liquid like) 
exciton gas and the high density system of interacting (and strongly 
overlapping) excitons in quasi-1D semiconductor (GaAs) QWR systems.
This exciton gas to a strongly overlapping and highly correlated electron-hole
system crossover with increasing electron-hole density can be thought of 
as a transition from an insulating exciton gas to a conducting 
electron-hole plasma (EHP), the Mott transition.
A typical feature of this Mott transition, 
observed in higher dimensional (2D,3D) optical experiments,
is the development of optical gain in the absorption spectra
where the absorption coefficient becomes
negative (gain region) in some frequency range.
One of the questions addressed in this paper is whether such an 
optical gain region exists in 1D photoexcited QWR systems.
In this paper we consider
the formation, stability, and optical properties of one-dimensional
excitons from low to high carrier densities in 
semiconductor QWR under photoexcitation conditions (i.e. equal electron 
and hole densities), a problem which has attracted a
great deal of both theoretical
\cite{glutsch,wang00,exciton_nojima_dim,exciton_nojima,exciton_simple,exciton_coulomb_rossi,ogawa91_optical,exciton_qwr,wf_qwr,exciton_papers,benhu,bgr_hwang,stopa,review_haug} and
experimental \cite{exciton_exp_amb,exciton_exp_weg,exciton_laser,bgr_exp}
attention in these years. Consistent with recent interest one of the central
issues we focus on is the density-induced exciton gas to EHP Mott transition
in 1D QWR systems and its experimental signature.

The motivation of our work arises from recent experimental studies of the
photoluminescence spectra of 1D GaAs/Al$_{1-x}$Ga$_{x}$As semiconductor QWR 
systems~\cite{exciton_exp_amb,exciton_exp_weg,exciton_laser,bgr_exp}.
The experimentalists use strong lasers to pump photons into the QWR 
systems, exciting electrons from the filled valence band into the 
empty conduction band
and/or the exciton levels, and observe the spectrum of the subsequently
emitted light coming from the eventual
recombination of the excited electrons and the holes created in the 
valence band. 
The photoluminescence spectrum is proportional to the exciton/EHP optical
oscillator strength, which, at first sight (i.e. without incorporating the
Sommerfeld factor effect associated with the electron-hole
Coulomb interaction), is expected to have
an $\omega^{-1/2}$ singularity at the band gap energy 
due to the $E^{-1/2}$ divergence of the 1D electron
density of states~\cite{review_haug} at band edge. However, this 1D plasma
band edge singularity is known to be strongly suppressed by the 
excitonic Coulomb correlation effect~\cite{exciton_coulomb_rossi} so that
the main peak observed in the experimental photoluminescence spectra
should result from 
the excitonic effect rather than the band-edge singularity of 
the noninteracting electron-hole plasma. 
The most striking experimental observation in the recent
~\cite{exciton_exp_amb,exciton_exp_weg,exciton_laser,bgr_exp} experimental
studies of photoexcited QWR systems has been the finding
\cite{exciton_exp_amb,exciton_exp_weg} that
the exciton peak seems to be at an almost constant energy independent of 
the carrier density, i.e. independent of the laser pumping power.
Thus the exciton peak seems to remain well-defined (and unshifted in energy)
all the way from very low to very high photoexcitation density 
($\sim 3\times 10^6$ cm$^{-1}$ \cite{exciton_exp_amb}) without
any distinct signature of 
the expected insulator(exciton)-to-metal(EHP) 
Mott transition and the associated optical gain.
The constancy of the exciton energy could, in principle, arise from an
almost exact cancelation between the exchange-correlation induced
shrinking of the nominal band gap, the so-called band gap renormalization 
(BGR), and the reduction
of the exciton binding energy (with respect to the 
bottom of the renormalized band edge) due to the screening induced
softening of the Coulomb interaction~\cite{wang00}.
Such an accidental cancelation between two distinct physical mechanisms
(namely, BGR and screening suppression of exciton energy) over a wide range 
of photoexcitation density needs to be theoretically established in a 
compelling way~\cite{wang00}. In addition, 
combining this accidental cancelation explanation with the experimental
fact of very a high Mott density (not yet seen experimentally)
one may conclude that 
the BGR of 1D electron-hole (e-h) system should be very weak in the high
density situation, which is not consistent with the theoretical calculations 
upto now~\cite{bgr_exp,benhu,bgr_hwang}. In particular, one must understand 
why there is no characteristic signature of the EHP in the luminescence 
spectra even at very high photoexcitation densities. One must be able to answer 
the question: where has the Mott transition gone ?
On this issue, an important and unresolved problem for the
photoluminescence experiment is that
there is no reliable and direct way of estimating the
photoexcited electron-hole density in such highly pumped QWR systems. 
The theoretical basis of the density estimation methods in the literature 
\cite{exciton_exp_amb}, such as from the lineshape analysis
of the spectrum,
is usually not self-consistent and not 
appropriate in such high density strongly laser-pumped systems.
Although we feel that the precise carrier density of the photoexcited 
QWR systems may not be known accurately, this issue does not pose any 
fundamental problem for our theory where the EHP density $n=n_e=n_h$ 
is an input parameter. The problem arises only in trying a direct 
quantitative comparison with experiments
\cite{exciton_exp_amb,exciton_exp_weg,exciton_laser}.

From the theoretical point of view, the full many-body calculation in 
a high electron-hole density semiconductor system is complicated 
and has not been attempted before except for our own short letter published
last year~\cite{wang00}.
The exciton mode is a solution of the Bethe-Salpeter equation (BSE) for 
the interaction vertex which, in the many particle situation of interest
to us, should include self-energy and dynamical screening correlations. 
A complete or exact solution of the BSE is only possible in the dilute 
exciton limit when it reduces to a simple hydrogenic 
electron-hole bound state Schr$\ddot{\rm{o}}$dinger equation. Our 
interest in this paper is in the many-particle "exciton" state in 
the photoexcited semiconductor QWR system where self-energy correlations 
of simple electron or hole states and dynamical screening of the 
electron-hole Coulomb interaction vertex are important. 
A model of an electron gas with a single hole in a wire~\cite{stopa} 
is not appropriate in our problem because a bound state always exists in 
any attractive potential in 1D systems,
which will trivially provide an overestimate of the Mott density. 
We emphasize that both the 
quasi-particle self-energy
and the dynamical screening of the electron-hole interaction vertex 
should be included properly (i.e. consistent with each other in a 
conserving approximation) in the BSE
to obtain the correct description of the Mott transition.
With the exception of our own short earlier report~\cite{wang00}
most other theoretical calculations use the
static (Hartree-Fock) approximation or the quasi-static 
approximation~\cite{review_haug} to the self-energy 
and a statically screened interaction vertex to solve the many-particle
BSE and obtain the optical absorption/gain spectra. In these simpler 
approximations, where dynamical screening effects are neglected in 
an uncontrolled way,
the dominant excitonic peak has a weak red-shift (a few meV decrease) 
with increasing density upto a Mott density, $n_c$, above which the excitonic
peak completely disappears and the spectrum shows a shallow (and weak) 
gain region very similar to the behavior of 
the noninteracting EHP
\cite{exciton_nojima_dim,exciton_coulomb_rossi,exciton_papers,stopa}.
Including the many-body dynamical screening~\cite{wang00} in the 
Coulomb interaction, 
the exciton peak stays essentially constant in energy (for $n<n_c$) 
and exhibits a pronounced
gain spectrum (for $n>n_c$), stronger than the quasi-static
results. But the predicted Mott densities in the above theories
($n_c\sim8\times 10^4$ - $8\times 10^5$ cm$^{-1}$) are all below the 
experimentally estimated value ($n_c>3\times 10^6$ cm$^{-1}$) --- in fact,
it is not clear if experimentally the transition to the EHP has ever been
observed even at the highest photoexcitation densities. 
It is in general hard to include the many-body effects appropriately
in a calculationally
tractable model over such a wide range of density
(over at least 4 orders of magnitude in $n$), from the weak coupling
dilute exciton gas system to the strong coupling EHP regime. 

In this paper, starting with  
the realistic Coulomb interaction in 1D T-shaped 
QWR systems, we first evaluate the single particle 
self-energy for both electrons and
holes in the dynamical plasmon pole approximation (PPA) within the so-called
GW scheme (i.e. in the leading order dynamically screened interaction)
to obtain the electron and hole renormalized Green's function. This 
self-energy calculation which by itself does not contain any direct 
excitonic effects, gives us the BGR or the reduction of the nominal 
band gap due to exchange-correlation.
For comparison, we also calculate the BGR obtained
by the quasi-static calculation in both static random phase approximation (RPA)
and static PPA in this paper. We then derive
analytically the effective electron-hole (e-h) interaction vertex, 
$V_{eff}(k,\omega)$, which includes consistently the
electron-hole-plasmon coupling with the external photons 
within our dynamical GW-RPA-PPA approximation scheme.
We use two different methods to study the excitonic 
properties: one is a variational approximation on
an effective exciton Hamiltonian~\cite{review_haug}, which depends on
the carrier density; the other technique
is to solve the dynamical BSE
by treating both self-energy renormalization and
vertex correction (arising from the Coulomb interaction) on an equal footing
(within our plasmon pole approximation scheme),
obtaining the optical absorption spectra.
Both calculations are carried out over a wide range of e-h density 
from $n=10^2$ cm$^{-1}$ to $n=10^6$ cm$^{-1}$ at
finite temperatures under the 
quasi-equilibrium condition, i.e. the e-h density is assumed to be a constant 
parameter for each density calculation (and $n=n_e=n_h$). 
While our dynamical BSE calculation includes exciton and EHP effects 
equivalently and is directly capable of providing the Mott density 
$n_c$ through the analysis of the absorption spectra, the variational 
exciton energy has to be compared with the BGR calculation in order 
to ascertain the Mott transition --- in particular, the merging of 
the effective variational exciton with the renormalized band edge 
is taken to be the signal for a Mott transition.
We find that the
absorption peak obtained from solving the dynamical BSE  
survives with very large broadening well above the critical density $n_c$ 
estimated from the variational approximation, and no optical
gain (negative absorption) regime shows up in the spectra even at the 
highest e-h density. This implies the \textit{non-existence} of
Mott transition in 1D electron-hole systems.
This striking result may be physically understood as arising from 
the fact that
the quasi-particle picture underlying the conventional Fermi 
liquid model fails
in high density 1D systems due to strong inelastic 
scattering by plasmons, associated with the generic 
non-Fermi liquid behavior in 1D systems. In fact, in 1D systems there is
no conventional EHP because there are no single particle excitations 
in an interacting 1D systems. This non-existence of single 
particle excitations or quasi-particles also leads to a  
breakdown of the conventional 
exciton picture --- a quasi-electron and a quasi-hole bound pair. 
We will discuss this point in more details later in this paper.

This paper is organized as follows: in Sec. II we present and discuss the 
theory we use in various parts of our calculations, e.g. 
the realistic Coulomb interaction in 1D T-shaped QWR system, the 
single particle self-energy calculation in the single loop PPA-GW 
approximation, the different approximation schemes used for screening
the long-ranged Coulomb interaction, 
the dynamical Bethe-Salpeter 
equation approximations in our theory, and the effective exciton 
Hamiltonian used in the variational calculation. In Sec. III we 
show our results for the density dependent exciton energy in the 
variational method and for
the excitonic optical properties from the solution of BSE. 
In Sec. IV we conclude with a
discussion and a summary of our results.
%%%%%%%%%%%%%%%%%%%%%%%%%%%%%%%%%%%%%%%%%%%%%%%%%%%%%%%
\section{Theory}
We use the two-band (one conduction band and one 
valence band) model to study the 1D electron-hole system, neglecting
higher subbands and the degenerate valence bands. We also 
consider the photoexcited 
quasi-equilibrium regime where the e-h density is assumed to be a constant
(in time) so that the Hamiltonian of such a 1D electron-hole system 
can be expressed as 
(in the effective mass approximation and assuming purely parabolic band
dispersion; we take $\hbar=1$ throughout):
\begin{eqnarray}
H&=&\sum_k\left[\left(E^0_g+\frac{k^2}{2m_e}c^\dagger_k c_k+
\frac{k^2}{2m_h}d^\dagger_k d_k\right)\right]
\nonumber\\
&+&\frac{1}{2L}\sum_{k,k',q}\left[V_{c,ee}(q)c^\dagger_{k-q}c^\dagger_{k'+q}
c_{k'}c_{k}+V_{c,hh}(q)d^\dagger_{k-q}d^\dagger_{k'+q}d_{k'}d_{k}
-2V_{c,eh}(q)c^\dagger_{k-q}c_{k}d^\dagger_{k'+q}d_{k'}\right],
\end{eqnarray}
where $c_k(c_k^\dagger)$ and $d_k(d_k^\dagger)$ are the 
annihilation(creation) operators for conduction band electrons and valence
band holes respectively (we will not explicitly show the spin index 
in summations throughout this paper although spin is included in our 
calculations), 
and $m_{e/h}$ are the electron/hole effective masses.
$E^0_g$ is the direct band gap between the top of the valence band and 
the bottom of the conduction band, taken to be 1550 meV for the 
GaAs/Al$_{1-x}$Ga$_x$As QWR system in all our calculations.  
There are three different Coulomb interactions
entering the Hamiltonian: electron-electron ($V_{c,ee}(q)$), hole-hole
($V_{c,hh}(q)$), and electron-hole ($V_{c,eh}(q)$) interactions.
The first two give rise to the electron and hole quasi-particle self-energies
and the other one,
the electron-hole interaction, produces the exciton bound state.
One should note that if we neglect the self-energy correction and also 
dynamical screening effect
(i.e. the low density limit of a Wannier exciton), the Hamiltonian of
Eq. (1) leads to a 1D hydrogen atom problem~\cite{manhan} for the Wannier
exciton, which in 1D
always has a bound excitonic state even for an arbitrarily weak 
electron-hole (attractive) interaction. Using a model of 
an electron gas with a single hole therefore gives rises
a very high Mott
density estimate (even if $V_{c,eh}$ is statically screened), which is 
a reflection of this 1D bound state property. We
address both the many-body self-energy effect and the electron-hole excitonic
binding effect on an equal footing in the theory, which we
accomplish by using the dynamical Bethe-Salpeter equation 
as described below.
%%%%%%%%%%%%%%%%%%
\subsection{Coulomb interaction in QWR}
The realistic (bare) Coulomb interaction in 1D QWR is obtained by taking the
expectation value of the 3D Coulomb interaction over the electron
wave function along the transverse directions ($y$ and $z$ axes, see the
inset of Fig. 1) of
the wire. After Fourier transformation along the 1D wire direction ($x$),
we have~\cite{exciton_qwr,benhu,Lai} for the Coulomb interaction 
matrix element:
\begin{eqnarray}
V_{c,ij}(q)&=&\frac{\rm{e}^2}{\varepsilon_0}\int_{-\infty}^{\infty}\it{}dy
\,dy'\int_{-\infty}^{\infty}dz\,dz'
\int_{-\infty}^{\infty}dx
\frac{e^{-iqx}|\phi_i(y,z)|^{\rm 2\it}|\phi_j(y',z')|^{\rm 2\it}}
{\sqrt{x^{\rm 2\it}+(y-y')^{\rm 2\it}+(z-z')^{\rm 2\it}}}
\nonumber\\
&=&\frac{2\rm{e}^2}{\varepsilon_0}\int_{-\infty}^{\infty}\it{}dy
\,dy'\int_{-\infty}^{\infty}dz\,dz'
|\phi_i(y,z)|^2|\phi_j(y',z')|^2
K_0(q\sqrt{(y-y')^2+(z-z')^2}),
\end{eqnarray}
where $\phi_i(y,z)$ is the QWR confinement wavefunction for 
the lowest eigenstate 
of electrons ($i=e$) or holes ($i=h$). Their exact forms 
depend on the geometry and the detailed nature of confinement for 
the QWR system.
$K_0(x)$ is the zeroth-order
modified Bessel function of the second kind~\cite{Lai} 
which diverges logarithmically
when $x$ goes to zero (i.e. in the long wavelength limit).

Following the experimental system of Ref.~\cite{exciton_exp_weg},
we use T-shaped QWR parameters to numerically calculate 
the 1D Coulomb interaction via Eq. (2). 
To simplify calculations (and also to have some analytical control) 
we use the following two approximations in evaluating the
wavefunction $\phi_i(y,z)$:
(i) we assume the confinement potential for both electrons and holes
to be infinitely deep, i.e. both electrons and holes are 
completely confined by the 2D T-shaped potential well, so that the wavefunctions
of electrons and holes are of the same form, independent 
of their effective mass difference. Consequently the three different
interactions ($V_{c,ee}$, $V_{c,hh}$, and $V_{c,eh}$) become
the same, denoted by $V_c$ throughout this paper. 
This simplifying approximation is justified by the detailed work of 
Ref.~\cite{glutsch},
as mentioned in the Introduction.
(ii) Instead of numerically solving the complicated
2D Schr$\ddot{\rm o}$dinger equation to get
the ground state single particle wavefunction~\cite{glutsch,exciton_qwr,wf_qwr}
(which is not the focus of our interest), we simply approximate $\phi(y,z)$
to be the product of two single-variable functions, $\xi(y)$ and $\psi(z)$
(i.e. $\phi(y,z)\sim\xi(y)\psi(z)$), and assume~\cite{Lai2} 
a simple and reasonable 
approximate model form for $\xi(y)$ and $\psi(z)$ through
the following exponential formulae: 
\begin{eqnarray}
\xi(y)&=&\frac{2^{3/4}}{W_y^{1/2}\pi^{1/4}}\,e^{-(2y/W_y)^2}
\\
\psi(z)&=&\frac{2^{5/2}z}{W_z^{3/2}}\,e^{-2z/W_z},
\end{eqnarray}
where $W_z$ and $W_y$ are the full-plane ($x-y$ plane) QW width
and the half-plane ($x-z$ plane) QW width respectively. Eqs. (3) and (4)
have the maximum electron/hole density at $y=0$, $z=W_z/2$, with three
branches of exponentially decaying density along $\pm y$ and $+z$ directions
(see the inset of Fig. 1). The exponential decaying
lengths or confinement sizes are $W_y/2$ and $W_z/2$ 
in $y$ and $z$ directions respectively, 
and thus in our model of the T-shaped QWR the effect of wire geometry on the 
Coulomb interaction is entirely contained in the effective "wire size" 
$W_y$ and $W_z$, which are the confinement parameters of our model.
This approximation greatly
simplifies the calculation of the realistic Coulomb interaction in Eq. (2)
and makes our BSE calculations tractable. We believe that our QWR 
confinement model, as defined in Eqs. (3) and (4), to be quite 
reasonable~\cite{Lai2}. 
For example, the exciton binding energy calculated in this approximation
is 18.2 meV for $W_y=W_z=7$ nm wire, very close to the quoted 
experimental value, 
17 meV, for the same wire size~\cite{exciton_exp_weg}. The small
over-estimate (about $7\%$) is expected because of the assumption of
infinite confinement energy and the strong $e^{-y^2}$ localization
of $\xi(y)$. In more accurate numerical treatments the confinement is
weaker than in our model, leading to a lower binding energy in the
QWR system.
In Fig. 1 we show the calculated $V_c(q)$ from Eqs. (2)-(4) for different 
wire sizes. We assume only one (the ground) electron and hole subband
in the conduction and valence band respectively.
%%%%%%%%%%%%%%%%%%
\subsection{Absorption spectra}
In order to study the excitonic effect on optical properties of 1D
photoexcited electron-hole systems in semiconductor QWR structures,
we calculate the dynamical (photon frequency dependent)
absorption coefficient, $\alpha(\omega)$, and refractive index, $n(\omega)$, 
which are related to the long wavelength
dielectric function, $\varepsilon(q\rightarrow 0,\omega)$, by the
following formula
\begin{equation}
n(\omega)+i\frac{c\alpha(\omega)}{2\omega}=\varepsilon(\omega)^{1/2},
\end{equation}
where $c$ is the vacuum light velocity.
The dynamical refractive index, $n(\omega)$, is therefore given in terms of
$\varepsilon(\omega)$ by,
\begin{equation}
n(\omega)=\sqrt{\frac{1}{2}\left[\rm{Re}\varepsilon(\omega)+
\left(\rm{Re}\varepsilon(\omega)^2
+\rm{Im}\varepsilon(\omega)^2\right)^{1/2}\right]}
\end{equation}
and the absorption coefficient, $\alpha(\omega)$, is given by
\begin{equation}
\alpha(\omega)=\frac{\omega\rm{Im}\varepsilon(\omega)}{n(\omega)c}.
\end{equation}
Using the linear response theory~\cite{review_haug,haug85},
the dielectric function
of the 1D e-h system is expressed as
\begin{equation}
\varepsilon(\omega)\simeq\varepsilon_\infty-\frac{4\pi\rm{e}^2}{\it{A}L}
\sum_{k,k'}r_{vc}(k)r^\ast_{vc}(k')G_{q\rightarrow 0}(k,k',\omega),
\end{equation}
where the retarded pair Green's function, $G_q(k,k',\omega)$, is
\begin{equation}
G_q(k,k',\omega)=
-i\int_0^\infty  e^{i\omega t}
\left\langle\left[d_{-k}(t)c_{k+q}(t),
c^\dagger_{k'+q}(0)d^\dagger_{-k'}(0)\right]\right\rangle_0 dt,
\end{equation}
and $A=W_yW_z$ is the cross sectional area of the QWR. In these equations
$q$ is the center of mass momentum of
the exciton which is set to zero (and hence not shown explicitly) 
in all our calculations below. 
$r_{vc}(k)$ is the dipole
matrix element, which can be simplified in the
effective mass approximation~\cite{review_haug}:
\begin{equation}
|r_{vc}(k)|\simeq\frac{M(k)}{\sqrt{4mE^0_g}},
\end{equation} 
where the reduced mass $m=m_e m_h/(m_e+m_h)$ and
\begin{equation}
M(k)=\left(1+\frac{k^2}{2mE^0_g}\right)^{-1}.
\end{equation}
By introducing a new function
\begin{equation}
Q(k,\omega)=\sum_{k'}M(k')G(k,k',\omega),
\end{equation}
the dielectric function in Eq. (8) can be expressed to be
\begin{equation}
\varepsilon(\omega)\simeq\varepsilon_\infty-\frac{\pi\rm{e}^2}
{\it{A}LmE^{\rm{0}}_g}
\sum_{k}M(k)Q(k,\omega),
\end{equation}
and the dynamical function $Q(k,\omega)$, which is essentially a two-particle
Green's function, satisfies the Bethe-Salpeter equation described
below.
%%%%%%%%%%%%%%%%%%%%%%%%%%%%%%%%%%%%%%%%%%%%%%%%%%%%%%%%%%%
\subsection{Bethe-Salpeter equations}
For the results to be presented in this paper, the many-body exciton
is given by the so-called Bethe-Salpeter equation \cite{review_haug} for
the two-particle Green's function
shown diagrammatically in Fig. 2(a), which
corresponds to a rather complex set of two-component
(electrons
and holes) coupled non-linear integral
equations which must be solved self-consistently with the bare interaction
being the Coulomb interaction in the QWR 
geometry. These equations are notoriously difficult
to solve without making drastic approximations, and in fact have never before
been solved in the literature \textit{in any dimensions} (except
for our own short report earlier~\cite{wang00}).
In carrying out the full many-body dynamical calculations for 
BSE we are forced to make some approximations. 
Our most sophisticated approximation
uses the fully frequency dependent dynamically screened electron-hole Coulomb
interaction in the single plasmon-pole approximation, 
which has been shown to
be an excellent approximation~\cite{ppa}
to the full random phase approximation
(RPA, see Fig. 2(c)) for 1D QWR system.
For the self-energy correction we use the
single-loop GW diagram shown in Fig. 2(b) with the 
screened interaction given by
PPA. Ward identities then fix the
vertex correction, entering Fig. 2(a), to be the appropriate ladder integral
equation.

For convenience, we use the finite temperature imaginary time
Matsubara frequency
Green's function formalism in our analysis.
The bare electron-hole two-particle Green's 
function without any e-h interaction is
\begin{equation}
G^0(k,k',z,\Omega)=G_e(k,\Omega-z)G_h(-k,z)\delta_{k,k'},
\end{equation}
and it corresponds to the two separate Green's function lines of electron and 
hole in Fig. 2(a). For each particle line, we have 
\begin{equation}
G_i(k,z)=\frac{1}{z-\varepsilon_{i,k}-\Sigma_i(k,z)+\mu_i},\ \ (i=e,h)
\end{equation}
where $\varepsilon_{e,k}\equiv k^2/2m_e+E_g^0$ and 
$\varepsilon_{h,k}\equiv k^2/2m_h$ are the bare (noninteracting) 
band energies for electrons in the conduction band and for
holes in the valence band respectively; $\mu_i$ is the chemical potential 
and $\Sigma_i(k,z)$ is the self-energy (for a complex frequency, $z$),
which we will calculate later within GW approximation. In order to
avoid the multi-pole (and any possible branch cut)
structure in $G_i(k,z)$, we approximate $\Sigma_i(k,z)$
by the momentum-dependent band gap renormalization, $\Delta_i(k)$, which
is related to the self-energy through the self-consistent Dyson's
equation: $\Delta_{i}(k)=\Sigma_i(k,
\varepsilon_{i,k}+\Delta_{i}(k)-\mu_i)$, i.e. $\Delta_i(k)$ is the so-called
quasi-particle on-shell self-energy. However, $\Delta_{i}(k)$ 
can be well approximated~\cite{rice} by truncating
this equation at the first nontrivial order, i.e. 
$\Delta_{i}(k)=\Sigma_i(k,\varepsilon_{i,k}-\mu_i)$, which should be
reasonably valid in our calculations below.
Therefore we have the following electron/hole single pole Green's function,
\begin{equation}
G_i(k,z)\sim\frac{1}{z-\varepsilon_{i,k}-
\Delta_i(k)+\mu_i},
\end{equation}
for later calculations in this paper. 
The details of calculating the self-energy $\Sigma_i(k,z)$ within
the GW approximation are discussed in the Sec. II-D below.
 
The Bethe-Salter equation in Fig. 2(a) could be read as
(with $\beta=1/k_BT$, where $T$ is the temperature)
\begin{equation}
G(k,k',z,\Omega)=G^0(k,k',z,\Omega)\times\left(1+
\frac{1}{\beta}\sum_{k'',z}V_s(k-k'',z-z')G(k'',k',z',\Omega)\right).
\end{equation}
Putting Eqs. (14)-(16) into Eq. (17) we get
\begin{eqnarray}
&&
(\Omega-\varepsilon_{e,k}-\varepsilon_{h,-k}-
\Delta_e(k)-\Delta_h(-k)+\mu_e+\mu_h)G(k,k',z,\Omega)
\nonumber\\
&&=\left(G_e(k,\Omega-z)+G_h(-k,z)\right)\delta_{k,k'}\times
\left(1+\frac{1}{\beta}\sum_{k'',z'}V_s(k-k'',z-z')
G(k'',k',z',\Omega)\right).
\end{eqnarray}
This equation, however, is not of closed form and is difficult
to evaluate since it is a rather complex multidimensional singular
integral equation. We therefore have to use
an additional simplifying approximation first introduced by 
Shindo~\cite{review_haug,shindo,zimmerman}, where 
the two particle Green's function, $G(k,k',z,\Omega)$, is replaced
by a simple pair
Green's function $G(k,k',\Omega)$ (whose retarded function yields via Eq. (9)
directly the optical dielectric function):
\begin{eqnarray}
&&G(k,k',z,\Omega)\simeq
\nonumber\\
&&\frac{G_e(k,\Omega-z)+G_h(-k,z)}
{-\frac{1}{\beta}\sum_z(G_e(k,\Omega-z)+G_h(-k,z))}G(k,k',\Omega),
\end{eqnarray}
where
\begin{equation}
G(k,k',\Omega)\equiv -\frac{1}{\beta}\sum_z G(k,k',z,\Omega),
\end{equation}
and
\begin{eqnarray}
&&-\frac{1}{\beta}\sum_z(G_e(k,\Omega-z)+G_h(-k,z))=
\nonumber\\
&&1-n_e(\xi_{e,k})-n_h(\xi_{h,-k}).
\end{eqnarray}
Here $\xi_{i,k}\equiv \varepsilon_{i,k}+\Delta_{i}(k)$
and $n_{i}(\xi_{i,k})$ is the fermion momentum distribution 
function, $(e^{\beta(\rm{Re}\it\xi_{i,k}-\mu_i)}+1)^{-1}$, which keeps the 
electron and hole density constant by adjusting the chemical potential, $\mu_i$,
to satisfy the correct density constraint, $\int (dk/\pi)\,n_{i}(\xi_{i,k})=n$.
Note that the approximation defined by Eq. (19) follows from the exact 
BSE in a statically screened Coulomb interaction~\cite{shindo},
i.e. if the frequency dependence of the effective dynamically screened
interaction is neglected. We expect the Shindo approximation to be
a reasonable approximation in our dynamical 
calculation below, because
the dynamical screening effect contributes mostly to 
the correlation energy, whose real part
is dominated by the (static) Hartree-Fock exchange energy in the
high density region \cite{manhan} (while the imaginary part of the
correlation energy plays an important role in our calculations below).  
We have not been able to find a tractable way of solving the dynamical BSE 
without making the Shindo approximation.

Using Eqs. (15), (19)-(21) in Eq. (18), we then have the following
effective Bethe-Salpeter equation for the pair Green's function
$G(k,k',\omega)$ (after the analytical continuation
$\Omega\rightarrow \omega+i\delta-\mu_e-\mu_h$):
\begin{eqnarray}
&&G(k,k',\omega)=G^0(k,k',\omega)
\nonumber\\
&&\times\left(1-\sum_{k''}V_{eff}
(k'',k',\omega)G(k'',k',\omega)\delta_{ss''}\right),
\end{eqnarray}
where $G^0$ and the dynamically screened effective
electron-hole interaction, $V_{eff}$, are expressed as
\begin{equation}
G^0(k,k',\omega)=\frac{1-n_e(\xi_{e,k})-n_h(\xi_{h,-k})}
{\omega+i\delta-\varepsilon_{e,k}-\varepsilon_{h,-k}-\Delta(k,\omega)}
\delta_{k,k'},
\end{equation}
and
\begin{eqnarray}
V_{eff}(k,k',\omega)&=&
\left(\frac{1}{\beta}\right)^2\sum_{z,z'}\left[
\frac{G_e(k,\Omega-z)+G_h(-k,z)}{1-n_e(\xi_{e,k})-n_h(\xi_{h,-k})}
V_s(k-k',z-z')\right.
\nonumber\\ 
&\times&\left.
\frac{G_e(k',\Omega-z')+G_h(-k',z')}
{1-n_e(\xi_{e,k'})-n_h(\xi_{h,-k'})}\right]
_{\Omega=\omega-\mu_e-\mu_h+i\delta}.
\end{eqnarray}
The effective BGR, $\Delta(k,\omega)$, is given by
\begin{eqnarray}
\Delta(k,\omega)&=&\sum_{k'}\left[(1-n_e(\xi_{e,k'+q})-n_h(\xi_{h,-k'}))
\times V_{eff}(k,k',q,\omega)-V_c(k-k')\right]\delta_{ss'}
\nonumber\\
&=&-\sum_{k'}[n_e(\xi_{e,k'+q})+n_h(\xi_{h,-k'})]V_{eff}(k,k',\omega)
+\sum_{k'}[V_{eff}(k,k',q,\omega)-V_c(k-k')].
\end{eqnarray}
In Eq. (25) the self-energy term, $(n_e+n_h)V_{eff}$,
and the vertex correction, $V_{eff}-V_c$, are treated 
on an equal footing.  
$G^0(k,k',\omega)$ in Eq. (23) is the electron-hole 
pair Green's function with self-energy correction but without
electron-hole attractive interaction, which is now replaced by the dynamically
screened effective interaction, $V_{eff}(k,k',\omega)$, in the BSE, Eq. (22). 
If we neglect dynamical effects in $V_s(k,z)$
(as in the static or the quasi-static
approximation described below), then $V_{eff}(k,k',\omega)=V_s(k)$ according
to Eqs. (24) and
(21). In the following section, we will discuss the use of different screening
models to evaluate $V_{eff}(k,k',\omega)$ (and BGR, $\Delta_{e/h}(k)$, through
the screened GW approximation) in calculating the absorption spectrum
by solving the BSE.

Combining the Bethe-Salpeter equation, Eq. (22), for $G(k,k',\omega)$
with Eq. (12), we have the following equation for $Q(k,\omega)$:
\begin{equation}
Q(k,\omega)=Q_0(k,\omega)
\times\left(1-\frac{1}{M(k)}\sum_{k'}V_{eff}
(k,k',\omega)Q(k',\omega)\right),
\end{equation}
for $Q_0(k,\omega)\equiv \sum_{k'}M(k')G^0(k,k',\omega)$.
Once $Q(k,\omega)$ is obtained by solving the integral equation, Eq. (26),
which is also a BSE, it is straightforward to calculate the absorption
and gain spectra from the dielectric function, $\varepsilon(\omega)$, 
through Eq. (13).
%%%%%%%%%%%%%%%%%%%%%%%%%%%%%%%%%%%%%%%%%%%%%%%%%%%%%%%%%%%
\subsection{Self-energy, BGR and screening in QWR}
In order to solve 
Eqs. (22) to (26) for the Bethe-Salpeter equation, we have to use a screened
interaction, $V_s(k,z)$, in Eq. (24) to get $V_{eff}$ and also to get the
single particle self-energy, $\Sigma_i(k,z)$, in the Green's 
function of Eq. (16). In this section, we discuss and compare 
both the quasi-static approximation
and the dynamical (PPA) approximation in the screening calculation.
For convenience, 
we first discuss the self-energy part and then the screening effect.

In the GW approximation which is the leading-order self-energy in 
the screened interaction expansion, the self-energy is calculated 
in the single-loop
diagram composed of a noninteracting particle line and a screened 
interaction line (Fig. 2(b)). Using static screening in the 
interaction line (i.e. $V_s(k,z)=V_s(k,0)$), we get a screened 
exchange self-energy term only, and all higher order screening effects to the 
correlation energy are neglected. This approximation (named 
static approximation) is therefore too simplistic to give correct 
results~\cite{wang00}, although it has been extensively employed in excitonic
calculations because of its simplicity. An improvement to the 
static approximation is the quasi-static approximation~\cite{review_haug}, 
which neglects the recoil energy during the scattering process so 
that no dynamical frequency inside the screened interaction 
potential shows up. This approximation produces an extra constant Coulomb-hole 
term (the second term of Eq. (27)) in the self-energy in addition to 
the screened exchange self-energy of the static approximation, so 
that the full expression for the BGR in this quasi-static approximation 
becomes
\begin{eqnarray}
&&\Delta_{i}(k)=
\sum_{k'}\left[-V_s(k-k')n_i(\varepsilon_{i,k})+\frac{1}{2}(V_s(k')-V_c(k'))
\right],
\end{eqnarray}
where $V_s(k)\equiv V_s(k,\omega=0)=V_c(k)/\varepsilon(k,\omega=0)$ 
is the statically screened Coulomb
interaction, which could be analytically derived either from RPA
(using Eq. (29) below)~\cite{benhu}
or PPA (using Eq. (30) below)~\cite{ppa}. In our paper, the former 
is called quasi-static-RPA and the latter named quasi-static-PPA. 
Note that $\Delta_{i}(k)$ in Eq. (27) is pure real, i.e. without any 
imaginary part of the self-energy or
inelastic broadening effect, so that the quasi-particle assumption for
the Landau-Fermi liquid is completely satisfied in this approximation 
with an infinite quasi-particle life time.
It is well-known, however, that the quasi-particle assumption
breaks down in 1D (unlike in 2D or 3D)
electronic systems, with a generic non-Fermi liquid behavior
\cite{Review_Voit}. For the purpose of comparison
we still use this approximation to calculate the 1D optical 
properties in order to compare with the full dynamical calculation
results and to study the
quantitative validity of this widely-used  
quasi-static approximation both in the
higher dimensional systems~\cite{review_haug,haug85} and in the 1D system
\cite{exciton_nojima_dim,exciton_coulomb_rossi} in the literature. 
In Fig. 3(a) we show the conduction band energy,
$\xi^0_{e,k}-E_g^0=\varepsilon_{e,k}+\Delta_{e}(k)-E_g^0$, 
in the quasi-static PPA
for different electron densities. The band gap renormalization 
is almost a wavevector independent rigid shift in the quasi-static 
approximation.

For the self-energy $\Sigma_i(k,\omega)$ calculated in
the one-loop GW approximation with dynamically screened interaction, we have 
\begin{eqnarray}
\Sigma_i(k,z)&=&
-\frac{1}{\beta}\sum_{k',z'}V_s(k-k',z-z')G_i(k',z')
\nonumber \\
&=&-\frac{1}{\beta}\sum_{k',z'}\frac{V_c(k-k')}{\varepsilon(k-k',z-z')}
G_i(k',z'),
\end{eqnarray}
where we can use either RPA or PPA (which is an excellent approximation 
to RPA~\cite{ppa}) to calculate the dynamical dielectric function,
$\varepsilon(k,\omega)$. For zero temperature RPA, 
$\varepsilon(k,\omega)$ is obtained by including
the non-interacting polarizabilities of electrons 
($\Pi^0_e(k,\omega)$) and holes
($\Pi^0_h(k,\omega)$)~\cite{benhu}:
\begin{eqnarray}
\varepsilon(k,\omega)&=&1-V_c(k)\Pi^0_e(k,\omega)-V_c(k)\Pi^0_h(k,\omega)
\nonumber \\
&=&1-V_c(k)\sum_{i=e,h}\frac{m_i}{\pi k}\ln\left[
\frac{\omega^2-[(k^2/2m_i)-kv_{F,i}]^2}{\omega^2-[(k^2/2m_i)+kv_{F,i}]^2}
\right],
\end{eqnarray}
where $v_{F,e/h}$ is the (Fermi) 
velocity of electrons/holes at Fermi momentum in
the conduction/valence band. In this paper we will use
RPA in only calculating the quasi-static screening via Eq. (27) by
setting $\omega=0$ in Eq. (29),
not in the full dynamical BSE (Eq. (26)), 
because the pole structure (and branch cut properties) of 
screened interaction, $V_c(k)/\varepsilon(k,\omega)$, in the full dynamical 
RPA is too complicated
to deal with in the frequency summation of Eq. (24) and in the integral
equation of 
Eq. (26). In the dynamical PPA, however, the dielectric function  
$\varepsilon(k,\omega)$ is defined by the following expression
where screening is induced by a single (plasmon) 
pole satisfying the corresponding
$f-$sum rule~\cite{ppa}: 
\begin{equation}
\frac{1}{\varepsilon(k,\omega)}=
1+\frac{\omega_{pl}^2(k)}{(\omega+i\delta)^2-\omega^2_k},
\end{equation}
where $\omega_{pl}(q)=\sqrt{nV_c(q)q^2/m}$ is the 1D plasmon oscillator
strength and $\omega_q$ is the effective plasmon frequency given by
a simple formula~\cite{exciton_papers,review_haug,ppa}
\begin{equation}
\omega^2_q=\omega_{pl}^2(q)+\frac{nq^2}{m\kappa}+\frac{q^4}{4m^2},
\end{equation}
where $\kappa$ is the inverse screening length.
It has been shown that PPA is a very good approximation to RPA
in 1D systems, where plasmon excitations
dominate the single particle excitations~\cite{benhu,ppa}. The great virtue
of the single-pole PPA for our theory is that it makes 
our calculation of $V_{eff}$ in Eq. (24) 
tractable because the
integral equation in frequency becomes simple.
In the PPA the self-energy of electron ($i=e$) or hole ($i=h$)
can be expressed as a sum of the usual
exchange or Hartree-Fock energy, $\Sigma_i^{ex}(k)$, and the correlation energy,
$\Sigma_i^{cor}(k,\omega)$:
\begin{eqnarray}
\Sigma_i(k,\omega)&=&\Sigma_i^{ex}(k)+\Sigma_i^{cor}(k,\omega)
\nonumber \\
\Sigma_i^{ex}(k)&=&-\sum_{k'}V_c(k')n_i(\varepsilon_{i,k'})
\nonumber \\
\Sigma_i^{cor}(k,\omega)&=&\sum_{k'}\frac{V_c(k')\omega^2_{pl}(k')}
{2\omega_{k'}}\times
\nonumber\\
&&\hspace{-2cm}
\left[\frac{n_B(\omega_{k'})+n_i(\varepsilon_{i,k+k'})}
{\omega+\omega_{k'}-\varepsilon_{i,k'+k}-i\gamma}
+\frac{n_B(\omega_{k'})+1-n_i(\varepsilon_{i,k+k'})}
{\omega-\omega_{k'}-\varepsilon_{i,k'+k}+i\gamma}\right],
\end{eqnarray}
where $n_B(\omega_{k})$ is the bosonic momentum distribution function,
$(e^{\beta \omega_{k}}-1)^{-1}$, for the plasmons; $\gamma$ is a small
phenomenological damping term incorporating impurity scattering and
all other possible broadening process (see the discussion in Sec. III-B).
From Eq. (32) we see that the dynamical effect as well as the imaginary part
of $\Sigma_i(k,\omega)$ arises entirely from the correlation energy
(and is absent in the static (Hartree-Fock) or quasi-static theory). 
This will play an important role (which is crucial in 1D) 
in our following calculations.
Figs. 3(b) and 3(c) show the real and imaginary parts of electron energy,
$\xi_{e,k}-E_g^0=\varepsilon_{e,k}+
\Sigma_e(k,\varepsilon_{e,k}-\mu_e)-E_g^0$, taking into account the dynamical 
PPA self-energy renormalization.

Defining the on-shell self-energy to be $\Delta_{i}(k)\equiv
\Sigma_i(k,\varepsilon_{i,k}-\mu_e)$ where $i=e,h$,
the imaginary part of $\Delta_e(k)$ is proportional to the electron
inelastic-scattering rate~\cite{benhu} arising from electron-electron
interaction, which
is very small when $k$ is below some threshold
momentum $k_c$. For $k>k_c$ a new collective mode scattering 
channel opens up in which 
electrons lose energy by emitting plasmons. 
At zero temperature and for zero impurity scattering 
(clean system limit),
it can be shown that the inelastic scattering rate diverges as 
$(k-k_c)^{-1/2}$ when $k$ approaches $k_c$ from above in 1D~\cite{benhu}. 
Note that this divergence in Im$\Delta_{e/h}(k)$ 
also exists in the RPA calculation~\cite{benhu}, and is therefore
a characteristic of the interacting 1D system in the dynamical GW 
approximation, causing
a gap to open up at $k=k_c$ in the real part of the self-energy as shown
in Fig. 3(b). The existence of this gap in BGR (or the divergence in 
Im$\Delta_{e/h}(k)$)
reflects the breakdown of the quasi-particle picture
in the 1D electron system~\cite{Review_Voit} within the 
perturbative GW approximation.
An interacting 1D electron system is known 
to be better described
by the Luttinger liquid (LL) model than the Fermi liquid model
due to the strong plasmon scattering effect arising from the limited phase 
space in 1D. A Luttinger liquid, in contrast to a Fermi liquid, does not
have any discontinuity in its momentum distribution function, and
does not, therefore, have any true quasi-particles. The existence of a
Luttinger liquid is a purely nonperturbative effect of interaction 
and happens in 1D even for arbitrarily weak electron-electron interaction. 
We therefore cannot get a true Luttinger liquid within our perturbative 
GW approximation, but the opening of the gap in the real part of 
the self-energy (or equivalently the divergence in the imaginary 
part of the self-energy) is the perturbative signature of the 
breakdown of the Fermi liquid picture.  
At finite temperature
and for finite impurity scattering, the single particle properties 
calculated in the 1D Fermi liquid model via the dynamical GW approximation 
are similar to the results calculated in the Luttinger 
liquid theory. Therefore we believe that the strong inelastic 
scattering shown in Fig. 3(c) qualitatively reflects the LL character 
of 1D systems, and our self-energy calculation is qualitatively correct 
for our purpose of calculating excitonic optical properties.
Our inclusion of the strong inelastic scattering by plasmons catches
some essential aspects of 1D phase space restriction, which eventually
leads to the nonperturbative formation of a 1D Luttinger liquid, which 
is beyond the scope of this work.

We evaluate the effective interaction, $V_{eff}$ in Eq. (24), by 
using the same PPA approximation and obtain:
\begin{equation}
V_{eff}(k,k',\omega)=V_c(k-k')\left[1+\frac{1}{N_{eh}(k)}
\frac{1}{N_{eh}(k')}
\chi_{eh}(k,k',\omega)\right],
\end{equation}
where $N_{eh}(k)\equiv 1-n_e(\xi_{e,k})-n_h(\xi_{h,k})$
and $\chi_{eh}$ is given by the
the following complicated formulae containing eight different terms associated
with various dynamical processes in the 1D e-h system.
\begin{eqnarray}
&&
\chi_{eh}(k,k',\omega)=\frac{\omega_{pl}^2(k-k')}{2\omega_{k-k'}}\times
\nonumber\\ \nonumber\\
&&\left[
\frac{-(1+n_B(\omega_{k-k'}))n_e(\xi_{e,k})+
n_B(\omega_{k-k'})n_e(\xi_{e,k'})+n_e(\xi_{e,k})n_e(\xi_{e,k'})}
{\xi_{e,k}-\xi_{e,k'}-\omega_{k-k'}}\right.
\nonumber\\ \nonumber\\
&&
+\frac{-n_B(\omega_{k-k'})n_e(\xi_{e,k})+
(1+n_B(\omega_{k-k'}))n_e(\xi_{e,k'})-n_e(\xi_{e,k})n_e(\xi_{e,k'})}
{\xi_{e,k}-\xi_{e,k'}+\omega_{k-k'}}
\nonumber\\ \nonumber\\
&&
+\frac{-(1+n_B(\omega_{k-k'}))n_h(\xi_{h,-k})+
n_B(\omega_{k-k'})n_h(\xi_{h,-k'})+n_h(\xi_{h,-k})n_h(\xi_{h,-k'})}
{\xi_{h,-k}-\xi_{h,-k'}-\omega_{k-k'}}
\nonumber\\ \nonumber\\
&&
+\frac{-n_B(\omega_{k-k'})n_h(\xi_{h,-k})+
(1+n_B(\omega_{k-k'}))n_h(\xi_{h,-k'})-n_h(\xi_{h,-k})n_h(\xi_{h,-k'})}
{\xi_{h,-k}-\xi_{h,-k'}+\omega_{k-k'}}
\nonumber\\ \nonumber\\
&&
+\frac{n_e(\xi_{e,k})n_h(\xi_{h,-k'})+(1+n_B(\omega_{k-k'}))
(1-n_e(\xi_{e,k})-n_h(\xi_{h,-k'}))}
{\omega+i\gamma+\mu_e+\mu_h-\xi_{e,k}-\xi_{h,-k'}-\omega_{k-k'}}
\nonumber\\ \nonumber\\
&&
+\frac{-n_e(\xi_{e,k})n_h(\xi_{h,-k'})+n_B(\omega_{k-k'})
(1-n_e(\xi_{e,k})-n_h(\xi_{h,-k'}))}
{\omega+i\gamma+\mu_e+\mu_h-\xi_{e,k}-\xi_{h,-k'}+\omega_{k-k'}}
\nonumber\\ \nonumber\\
&&
+\frac{n_e(\xi_{e,k'})n_h(\xi_{h,-k})+(1+n_B(\omega_{k-k'}))
(1-n_e(\xi_{e,k'})-n_h(\xi_{h,-k}))}
{\omega+i\gamma+\mu_e+\mu_h-\xi_{e,k'}-\xi_{h,-k}-\omega_{k-k'}}
\nonumber\\ \nonumber\\
&&\left.
+\frac{-n_e(\xi_{e,k'})n_h(\xi_{h,-k})+n_B(\omega_{k-k'})
(1-n_e(\xi_{e,k'})-n_h(\xi_{h,-k}))}
{\omega+i\gamma+\mu_e+\mu_h-\xi_{e,k'}-\xi_{h,-k}+\omega_{k-k'}}
\right],
\end{eqnarray}
where we use the same phenomenological damping parameter, $\gamma$,
to broaden the resonant threshold energies in the denominators.
The first two terms in the bracket of Eq. (34) describe the coupling of
electron excitations with the plasmon, having the corresponding particle 
filling factors in the numerator and the resonance energy in the denominator.
The third and the fourth terms describe the same plasmon coupling 
process for the holes.
The first four terms are static and
$\omega$-independent in our approximation. The last 
four terms are dynamical and depend explicitly on $\omega$.
These last four dynamical terms describe processes which couple 
both electron and hole systems with the plasmon modes, and 
are extremely important in the dynamics of the photoexcited system.
%%%%%%%%%%%%%%%%%%%%%%%%%%%%%%%%%%%%%%%%%%%%%%%%%%%%%%%%%%%
\subsection{Effective Hamiltonian and variational method}
Before solving the full dynamical Bethe-Salpeter 
equation, it is instructive to study
the excitonic and the EHP effects \textit{separately} by treating the influence
of the EHP on
the excitonic states as a perturbation~\cite{review_haug,zimmerman}.
Using
an effective Hamiltonian derived from the Bethe-Salpeter equation,
we can variationally obtain the exciton ground state energy 
by minimizing the energy expectation value
through an 1$s$ exciton trial wavefunction.
The effective Hamiltonian treats the
EHP as a perturbative effect and is written as
$H_{pp'}(\omega_n)=H^0_{pp'}+H'_{pp'}(\omega_n)$, 
where 
\begin{equation}
H^0_{pp'}=\left(E_g^0+\frac{p^2}{2m}\right)\delta_{pp'}-V_c(p-p')
\end{equation}
is the Hamiltonian for the single electron-hole pair with an 
unscreened Coulomb interaction
(similar to a 1D hydrogen atom) 
and the perturbation $H'$ is
\begin{eqnarray}
H'_{pp'}(\omega_n)&=&\Delta(p,\omega_n)\delta_{pp'}+V_c(p-p') \nonumber\\
&&-(1-f_e(\xi_{e,p})-f_h(\xi_{h,-p}))V_{eff}(p,p',\omega_n),
\end{eqnarray}
for the $n$th eigenstate of energy $\omega_n$.
Here we can explicitly see the physical meaning of $\Delta(p,\omega)$ and 
$V_{eff}(p,p',\omega)$ analytically derived in Eqs. (24) and (25). 
We expect that the wave function of
the exciton satisfies the corresponding Schr$\ddot{\rm{o}}$dinger's
equation in the low density limit, where the screening effect is negligible.
Thus this exciton effective Hamiltonian approach may be a reasonable
approximation to calculate exciton energies and wavefunctions.

For the exciton trial wavefunction, $\phi_n(p)$, in the momentum space,
we use the two parameter variational wavefunction first introduced by
Nojima~\cite{exciton_nojima_dim,exciton_nojima}
to express the 1D exciton ground state as
\begin{equation}
\phi_0(p)=\sqrt{\frac{2\sigma\lambda}{K_1(2\sigma)}}\frac{K_1(\sigma
\sqrt{\lambda^2 p^2+1})}{\sqrt{\lambda^2 p^2+1}},
\end{equation}
where $\lambda$ and $\sigma$ are two independent (positive)
variational parameters in our calculation. $K_1(x)$ is
the first order modified Bessel function of the second kind.
This variational bound state wave function has the following 
form in the real space,
\begin{equation}
\phi_0(x)=\frac{\exp\left[-\sqrt{(x/\lambda)^2+\sigma^2}\right]}
{\sqrt{2\sigma\lambda K_1(2\sigma)}},
\end{equation} 
where one can see that the variational parameter $\lambda$ 
represents the exciton radius
and $\sigma$ smoothens or broadens 
the center of mass wavefunction at $x=0$.
We do not study the first excited 
state wave function, $\phi_1(p)$, in this paper because it is not 
particularly relevant
to the Mott transition process we are interested in, 
although the variational technique can be adapted to study excited 
excitonic states~\cite{exciton_nojima}.
%%%%%%%%%%%%%%%%%%%%%%%%%%%%%%%%%%%%%%%%%%%%%%%%%%%%%%%%
\section{Results}
We first show the variational results 
because conceptually this is the simplest
approach since it is based on an effective single particle Hamiltonian. 
We obtain the BGR and the exciton
binding energy by variational method
in both the quasi-static approximation and the dynamically 
screened GW approximations for
various photoexcited carrier (e-h) densities. 
The crossover between the exciton energy and the
BGR gives us an estimated  
Mott transition critical density, $n_c$,
where the exciton bound state
ceases to exist, and an insulator-to-metal transition occurs.
The idea here is that at $n_c$ the exciton merges with the e-h continuum and is no longer a stable bound state.
Finally we carry out the full Bethe-Salpeter integral equation solution by a
matrix inversion method and obtain the absorption spectra and refractive index
in a large range of e-h density (from $10^2$ to $10^6$ cm$^{-1}$) to compare
with the variational effective Hamiltonian results. 
Details are described below.
%%%%%%%%%%%%%%%%%%%%%%%%%%%%
\subsection{Effective Hamiltonian result}
In Fig. 4(a), we show the calculated
density dependence of the exciton ground state energy variationally
obtained from the effective Hamiltonian method and 
the BGR ($\Delta_e(0)+\Delta_h(0)$) calculated
in both the quasi-static approximation and the dynamically screened
GW approximation as described
in the section II-D. Both RPA and PPA are used in the quasi-static
calculation (Eq. (27)) for comparison whereas the full dynamical 
calculations are done only in PPA. The intersection between the
exciton energy (dashed lines) and the BGR (solid lines) indicates the Mott
transition, where the exciton merges with the band continuum and the system
has a phase transition from an insulating exciton gas to a conducting EHP.
Note that the variational method introduced 
in Sec. II-E loses its accuracy near the Mott
density (and becomes essentially meaningless for $n>n_c$),
because the variational energy expectation value,
$E(\lambda,\sigma)\equiv\langle\phi_0(\lambda,\sigma)
|H^0+H'(E(\lambda,\sigma))|\phi_0(\lambda,\sigma)\rangle$, has a very flat
minimum region in the $\lambda-\sigma$ space around $n\approx n_c$, 
i.e. the exciton wave function is highly
broadened, so that its minimum energy is hard to determine in such
perturbation-based variational method. In Fig. 4(b) we show the 
variationally calculated trial exciton 1$s$ (ground state)
wavefunction, $\phi_0(x)$, for different exciton densities. The exciton
density dependences of variational parameters, $\lambda$ and $\sigma$,
are also shown in the inset of Fig. 4(b). The sharp divergences of 
$\lambda$ and $\sigma$ at $n_c\sim 2\times 10^5$ cm$^{-1}$
indicate the delocalization of exciton ground
state wave function, a signal of exciton-to-EHP Mott transition.
In Fig. 4(a) we terminate the variationally
calculated exciton line (dashed) at $n=2\times 10^5$ cm$^{-1}$ and
use the dotted line to
represent the peak position of the absorption spectra obtained from
solving the full dynamically screened BSE (discussed below) to 
continue the exciton line
to higher densities upto $6\times 10^5$ cm$^{-1}$. 

We can make the following comments about the results shown in Fig. 4(a): 
(i) For density
below $10^4$ cm$^{-1}$ the exciton energy has only a few meV 
density-dependent red-shift
in the quasi-static-RPA/PPA approximations and almost no shift 
(less than 0.5 meV) in the dynamical screening approximation. This shows
the almost complete cancelation between the exchange-correlation induced 
BGR (a density-dependent shift) and the blue-shift of the
exciton energy (due to screening) over a wide range of density.
On the other hand, using the static screening (i.e. exchange energy only) 
approximation in the same calculation does not lead to 
this cancelation~\cite{wang00}, showing that the experimentally observed
constancy of the exciton energy as a function
of the photoexcited e-h density is a dynamical effect, which may not manifest
itself in simpler approximations.
(ii) For an e-h density higher than $10^4$ cm$^{-1}$, the exciton
energy in the quasi-static-RPA has a rather 
large red-shift until it merges with the BGR line
smoothly at $n_c\sim 6\times 10^4$ cm$^{-1}$, indicating a rather
low density Mott transition in this system. On the other hand, the exciton
energies calculated in both the
quasi-static-PPA and the full dynamical PPA are almost constant upto
$n=n_c\sim 3\times 10^5$ cm$^{-1}$, where the band continuum meets the 
exciton energy.
(iii) In the full dynamical results obtained by 
solving the dynamical BSE directly, 
the excitonic absorption peak (dotted line)
seems to survive even for densities higher than the "critical density", $n_c$,
at which the calculated exciton energy crosses the band continuum
BGR line. This shows that there must be reasonably strong hybridization 
between the exciton and the EHP in
the dynamical BSE (note that the dotted line in Fig. 4(a) is 
not from the variational calculation, but is obtained from the BSE 
solution), so that 
the effective BGR, including excitonic effects in the BSE, is actually
less than the result we calculate 
from Eq. (32) by adding the electron and hole self-energies without 
incorporating exciton effects. This also demonstrates
the limitation of the quasi-static approximation and confirms the
necessity of the full dynamical BSE calculation in the high density 
1D e-h system. 

%%%%%%%%%%%%%%%%%%%%%%%%%%%%%%%%%%%%%%%%%%%%%%%%%%%%%%%
\subsection{Dynamical Bethe-Salpeter equation result}
In Fig. 5, we show our calculated absorption and gain spectra
by solving the full Bethe-Salpeter (integral) equation in the quasi-static
and the full dynamical screening approximations for $W_y=W_z=7$ nm wire at
a low temperature of $T=10$ K. The integral equation for the
two-particle Green's function, Eq. (26) (or equivalently Eq. (22)), 
is solved by the matrix inversion method with maximum momentum upto
$k_{max}=(\pi/2)\times 10^8$ cm$^{-1} (=100k_F$ for $n=10^6$ cm$^{-1}$).
The poles of the dynamical screened interaction, $V_{eff}$, in Eqs. (33) 
and (34), together with the logarithmic
singularity of the 1D Coulomb interaction in the long wavelength limit,
produce a multi-singular
kernel with multiple momentum-dependent singularities
which have never been solved before in the literature (except for our
earlier work~\cite{wang00}), because the usual 
singularity-removal method is ineffective here~\cite{review_haug,haug85}.
In our calculations presented in this paper, we use
a rather large matrix ($1500\times1500$ in a Gaussian quadrature for 
$|k|\leq k_{max}$) in the
matrix inversion method in order to get good overall accuracy, i.e.
the same calculations using even larger ($2000\times2000$) matrix 
size (which is extremely time-consuming and not shown here) do not
show any significant difference (within 10\%) in the whole 
absorption (and refractive index) spectrum from the results 
we present in Figs. 5-7.
The broadening $\gamma$ used in our calculation is a phenomenological 
parameter which simulates in a simple manner the effects of all 
possible scattering and broadening processes not explicitly included 
in our theory. These are, for example, impurity and defect scattering, 
inhomogeneities in the system (e.g. fluctuations in the wire width), 
broadening associated with optical excitation process itself, and 
phonon scattering. We mention that inelastic plasmon scattering life 
time effects are explicitly included in our theory. Note that $\gamma$ 
should be small compared with the bare excitonic binding energy 
($\sim 10-20$ meV in GaAs semiconductor QWR systems), and as long 
as $\gamma$ is small, its precise value has no qualitative effect 
on our conclusions and results. We typically choose $\gamma=0.5$ 
meV in our calculations.

In Fig. 5(c) and 5(d) we show the absorption spectra in the dynamical PPA 
for two different values of impurity scattering $\gamma$ (different by
a factor of 2.5) to show that $\gamma$ does not affect the qualitative 
behavior of the spectra, but does control the linewidths of the absorption
peaks as one would expect.
Some important features of the optical spectra (calculated by solving
the full BSE) shown in Fig. 5 are:
(i) there are generally two absorption peaks in the low density
($n<10^4$ cm$^{-1}$) spectra of all the three approximations,
one is the exciton ground state (1$s$) peak at about 
1532 meV and the other one is the exciton first excited state (2$s$) at,
for example, 1547.5 meV for $n=10^2$ cm$^{-1}$
(this peak is off the plot region in Fig. 5(a)). 
Note that this low density spectrum is almost the same in all the three
different approximations, showing that the dynamical effect is not important 
in the low density region.
(ii) When the density increases but is still less than $10^4$ cm$^{-1}$,
the exciton peak does not shift much ($<2$ meV) with
increasing carrier density in all approximations,
indicating the constancy of the exciton energy. (iii) At higher densities
($10^4$ cm$^{-1}$ $<n<$ $10^5$ cm$^{-1}$),
however, the quasi-static-RPA result (Fig. 5(a)) shows some additional
red-shift in the excitonic peak, consistent with the result
shown in Fig. 4(a) which is obtained from the variational method. 
On the other hand, the excitonic peak positions in the quasi-static-PPA 
and in the full dynamical approximation
are almost (density independent) constants in this region. A significant
difference between the quasi-static-PPA and the dynamical 
calculation results, however, is that
the exciton peak of the quasi-static-PPA results (Fig. 5(b)) has an almost
constant oscillator strength, while the oscillator strength of the peak 
in the full dynamical calculation results 
(Fig. 5(c)) decays at high density to about one-third of its low density value.
(iv) For $n>10^5$ cm$^{-1}$, both quasi-static-RPA and quasi-static-PPA
results show negative absorption (gain) for the photon frequency below
some critical value, $\omega_c$, while
the full dynamical result is still positive (i.e. no gain) with a weaker 
broadened peak
upto $n\approx 6\times10^5$ cm$^{-1}$ or higher. In other words, we do not 
explicitly find the expected exciton (insulator) to 
plasma (metal) Mott transition
when both the self-energy and the screened interaction are included dynamically
in the full BSE theory upto a rather high e-h density.
We believe that this behavior arises from the strong plasmon 
scattering effects in 1D as discussed in Sec. IV of this paper.
(Such strong inelastic scattering was not included in our earlier short
report \cite{wang00}, leading to the appearance of a gain
in the high density spectra above the Mott density.)
In Fig. 6, we show the refractive index, $n(\omega)$, calculated by solving 
the Bethe-Salpeter equation in both the quasi-static-PPA
and the full dynamical approximation for different photoexcitation densities.
The calculated refractive indices in these two approximations  
are similar in structure.

In Fig. 7, we show the calculated absorption/gain spectra obtained in both the 
quasi-static-PPA (Fig. 7(a)) and the full dynamical 
calculations (Fig. 7(b)) for the same wire width, $W_y=W_z=7$ nm, but at a
higher temperature ($T=100$ K) for various densities.
We find that the higher temperature (100 K) 
low density ($n<10^4$ cm$^{-1}$) absorption spectrum is 
almost the same as the corresponding lower temperature ($T=10$ K) spectra in 
Figs. 5(b) and (c),
while in the higher density region ($n>10^4$ cm$^{-1}$) the high
temperature exciton absorption
peak of the full dynamical calculation has a smaller 
red-shift in energy with a much larger broadening
than the quasi-static-PPA result. The quasi-static-RPA result at such high 
temperature (not shown here) has an even larger 
red-shift and broadening. We mention
that the gain in the absorption spectra of the quasi-static 
calculation at the lower temperature (Fig. 5(b)) is flattened and 
almost disappears in the higher temperature (100K) calculation results.
%%%%%%%%%%%%%%%%%%%%%%%%%%%%%%%%%%%%%%%%%%%%%%%%%%%%%
\section{Conclusion}
In this paper, we theoretically study, using reasonably realistic 
Coulomb interaction, the excitonic optical properties 
of a 1D QWR system by solving the many-body Bethe-Salpeter equation
using a number of approximations, the most sophisticated one being 
a treatment of both the self-energy and the vertex function in the 
dynamically screened GW approximation. Our calculation is applied to the 
experimentally studied T-shaped 
GaAs-Al$_{x}$Ga$_{1-x}$As 1D QWR systems for various densities 
and temperatures. We calculate
the electron and hole self-energies in the one-loop 
GW approximation diagram using different screening approximations:
the quasi-static-RPA, the quasi-static-PPA, and the dynamical (PPA) 
approximation.  
The quasi-static approximations give an almost rigid shift (the BGR effect)
to the band energy (see Fig. 3(a)), and there is no imaginary part
of the self-energy, i.e.
the quasi-particle life time is infinite.
This approximation may work well in 2D and 3D systems but fails completely
in 1D systems, because unlike in the higher
dimensional systems, the electrons in 1D system suffer very strong inelastic
scattering effect by virtue of restricted phase space. Therefore
the validity of the quasi-static approximation applied to 1D systems, 
which has been extensively used
in many theoretical works~\cite{exciton_nojima_dim,exciton_coulomb_rossi,exciton_papers,higher_d,screening_multiband}, 
is doubtful. In the dynamical calculation
we find that the electron and hole band gap 
renormalization has a gap opening up in its real part
and a consequent divergent singularity in its 
imaginary part at $k=k_c$ (Figs. 3(b) and (c)), 
where the quasi-particle energy is transferred to the plasmon excitations due
to very strong inelastic scattering by 1D plasmons. Although this 
perturbative GW self-energy is "unphysical" due to the failure 
of the Fermi liquid model in the 1D system~\cite{Review_Voit}, it still
gives a rather good qualitative description of the single-particle 
and the collective mode 
properties (compared to the correct Luttinger liquid model), 
particularly at finite temperatures 
and for finite impurity scattering~\cite{benhu}.
Our results in Figs. 3(b)
and (c) reflect an important generic feature of 1D systems: the quasi-particle
excitation has a very short life time (in fact, it does not exist) 
if the excitation momentum is higher
than some value $k_c$. This 1D feature associated with Luttinger liquid
properties of 1D systems strongly affects the stability of 1D excitons,
the bound quasi-electron and quasi-hole pairs, as we can see from the 
calculated absorption and gain spectra (Fig. 5).  

In Fig. 5 we find that the 
quasi-static approximation, which excludes inelastic scattering, gives rise
to a negative absorption (gain) region in the highly
photoexcited system.
The existence of gain means that the exciton state is saturated 
(fully occupied),
and therefore manifests a spontaneous emission,
rather than absorption. On the other
hand, the overall positive absorption (no gain) spectrum found in the 
dynamical calculation (Fig. 5(c)) upto the highest density is caused by the 
large imaginary part of the electron/hole on-shell self-energy,
Im$\Delta_{e/h}(k)$ (see Fig. 3(c), and Eq. (32)), which is
proportional to the inelastic-scattering rate and results from the
energy scattering through plasmon channel. In other words, the 
excitons, composed of bound pairs of quasi-electrons and quasi-holes, are 
unstable due to strong inelastic scattering by 1D plasmon 
excitations in the high density region.
Consequently in the dynamical calculation, the exciton absorption peak 
is suppressed in strength and broadened in width as the photoexcitation
density increases leading to stronger plasmon scattering. The absorption 
spectrum does not exhibit a
negative (gain) region even in the high density regime because the 
quasi-particle EHP band continuum is 
so strongly inelastically scattered by plasmons that it is not a  
proper eigenstate (i.e. it decays) and is never saturated. 
The disappearance of the exciton line and the non-negativity 
in the absorption spectra (at the same time) in
our dynamical calculation suggest that there is
\textit{no} insulator (exciton) to metal (EHP) Mott transition in
1D systems, since both excitons and quasi-particles are
strongly inelastically scattered by plasmons leading to neither of them being
well-defined coherent states of the high density 1D system.   
This result is consistent with the well-known non-Fermi-liquid properties of
1D electronic systems, where the quasi-particle 
(and hence the exciton) picture fails. 
The quasi-static approximation, which ignores any plasmon effect and 
works well in 2D and 3D systems \cite{higher_d},
does not work in 1D systems because the 1D excitation spectrum is 
completely dominated by plasmons.

Another clue in support of the importance of plasmons in such high density 
1D e-h system comes from the temperature dependence
of the absorption spectra shown in Fig. 7. Our results
show that the high temperature ($T=100$ K) absorption peak 
in the dynamical calculation (Fig. 7(b)) is suppressed and broadened
so greatly that there is almost no spectral structure observed for
$n\geq 10^5$ cm$^{-1}$, while 
the high temperature quasi-static-PPA result (Fig. 7(a))
still has a rather strong peak at the same density. 
This is because the plasmon excitation occupancy, whose 
energy distribution function, $n_B(\omega_k)$, follows the
Bose-Einstein statistics, depends strongly on temperature, leading to 
a qualitative difference between the $T=10$ K and $T=100$ K results in 
the dynamical calculation, while such plasmon dynamics is not included
in the quasi-static calculation. This characteristic strong 
temperature dependence 
is also consistent with very recent experimental
results~\cite{exciton_laser}.

Based on our results and the discussion above,
we propose that a crossover from a low density (essentially
noninteracting) Fermi liquid to a high density interacting
non-Fermi liquid is occurring in the optical spectra of the 1D e-h system
as the photoexcitation density increases (see Fig. 5(c)). In the low density
limit, say $n\leq 10^2$ cm$^{-1}$, we have a dilute and noninteracting
exciton system, 
whose absorption spectrum is independent of the many-body screening 
approximations we use --- plasmons are just not that important in this regime. 
This shows that excitons in this situation 
are isolated quasi-electron
and quasi-hole pairs, reflecting the validity of the 
quasi-particle picture in the effective noninteracting Fermi
liquid model in the low density limit. In the higher 
density region, however, the 
plasmon effect on the quasi-particle self-energy
becomes important, because the band curvature at $k=k_F$ is
less for higher $k_F$ (i.e. higher density) and the relative importance of
collective mode excitations (plasmons) is then strongly enhanced
as in the Luttinger liquid model. Therefore
the oscillator strength of the exciton absorption peak is then reduced 
and broadened in
our dynamical calculation (Fig. 5(c)). When the density is roughly the nominal
"Mott transition" critical density, $n_c$, where the band continuum energy
equals the exciton energy (see Fig. 4(a)), the plasmon excitation becomes
so dominant
that both exciton and band continuum (EHP) states become unstable, showing 
a crossover to effectively non-Fermi liquid properties.
We therefore do not expect to see the real Mott transition from an excitonic 
insulator to an EHP metal in 1D highly photoexcited systems, in contrast
to the results of previous theories. For an electron-hole 
plasma without any backward 
scattering in the usual Luttinger liquid model 
(no band curvature at all, which is 
unrealistic in our case), we can prove 
that gain in the optical spectra does exist below the Fermi energy
at all densities with a
complicated power-law singularity at the Fermi
surface. Including the electron-hole attractive backward interaction (assuming
a short-ranged interaction as in the so-called $g-$ology formalism),
the electron-hole system most likely undergoes a charge density wave 
ground state transition
with a mass gap 
in the elementary plasmon excitation~\cite{ogawa95}. While this scenario
is consistent with our results, further work
for the excitonic effect at energy \textit{far below} the Fermi energy
is still needed, because the regular Luttinger liquid model cannot include 
band curvature in an appropriate way in order to study the Mott transition 
at an energy level around the band edge.

In reference to the experimental data,
we note that our results from solving the dynamically screened 
Bethe-Salpeter equation are in
excellent qualitative and quantitative agreement with the recent experimental
findings~\cite{exciton_exp_amb,exciton_exp_weg}.
In particular, the effective constancy of the exciton peak
as a function of the photoexcited carrier density as well as the possibility
of excitonic absorption well into the high density regime
(even for $n>6\times 10^5$ cm$^{-1}$) turn out to be characteristic
features of the full dynamical theory (but \textit{not} of the static and
the quasi-static approximation). A full dynamical self-consistent theory as
developed in this paper is thus needed for an understanding of the recent
experimental results. Moreover we find that in our theory,
the plasmon effect is crucial in the high density regime leading to
the non-existence of any observable Mott
transition in our calculation. This is consistent with 
recent experimental results~\cite{exciton_exp_amb,exciton_exp_weg}, which 
do not observe an actual Mott transition in the semiconductor QWR system
even in the high photoexcitation density ($\sim 3\times 10^6$ cm$^{-1}$)
regime. We emphasize that only our dynamical 
theory, and not the static or quasi-static approximation, is in
agreement with the experimental results. 

In summary, our main accomplishments reported in this paper are the following:
(i) the \textit{first} fully dynamical theory of a photoexcited electron-hole 
system in semiconductors which treats self-energy, vertex corrections, and 
dynamical screening in a self-consistent scheme within a realistic Coulomb 
interaction-based Bethe-Salpeter theory; (ii) a reasonable qualitative and 
quantitative agreement with the recent experimental observations of a 
constant (photoexcitation density-independent) excitonic absorption peak
in energy, which in our dynamical
theory arises from an approximate cancelation between the
self-energy and the vertex corrections in the Bethe-Salpeter equation; 
(iii) inclusion of the plasmon effect in the quasi-particle self-energy
calculation in our dynamical theory, leading to our theoretical proposal
that \textit{no} Mott transition should be observed in 1D electron-hole 
systems (at least in optical 
experiments) even at very high photoexcitation density, i.e. there should be
no optical gain region;
(iv) instead, we suggest an experimentally observable crossover
from a low density noninteracting Fermi liquid behavior 
(quasi-particle/exciton favored)
to a high density interacting non-Fermi liquid behavior 
(no stable quasi-particles and excitons).
A more precise and nonperturbative theoretical model for the high density
1D electron-hole system is needed for future study --- such a study 
should somehow incorporate both band curvature and Luttinger liquid 
behavior in analyzing the optical properties, although we believe 
that the qualitative features of such a theory are already 
contained in our work.

This work has been supported by the US-ONR and the US-ARO.
%%%%%%%%%%%%%%%%%%%%%%%%%%%%%%%%%%%%%%%%%%%%%%%%%%%%%%%%%%%%%%%%%%%%%%%%%

%%%%%%%%%%%%%%%%%%%%%%%%%%%%%%%%%%%%%%%%%%%%%%%%%%%%%%%%%%%%%%%%%%%%%%
\newpage
%--------------------------------
\begin{figure}
 \vbox to 6.1cm {\vss\hbox to 5.5cm
 {\hss\
   {\includegraphics{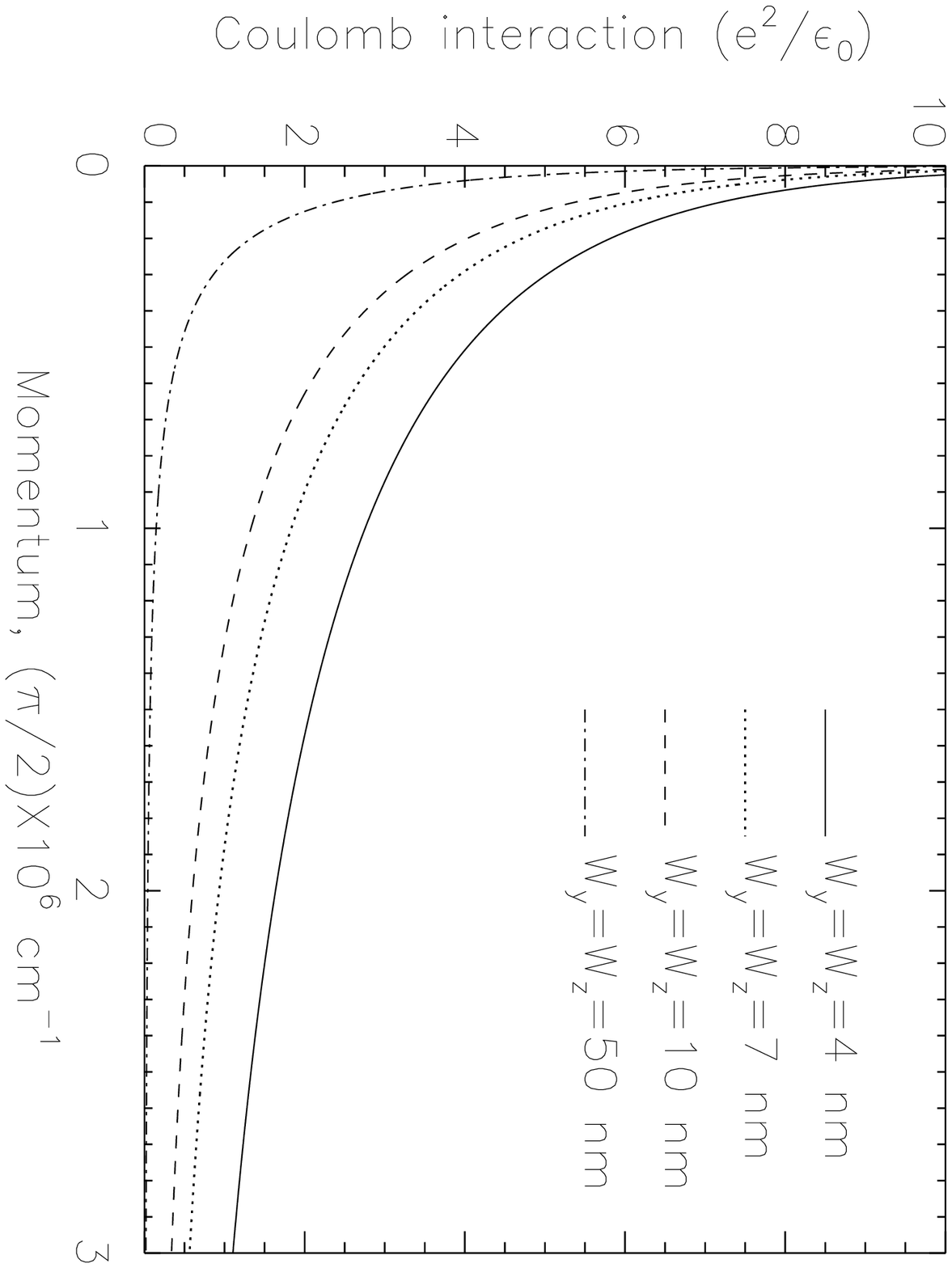}
   }
  \hss}
 }

 \vbox to 0cm {\vss\hbox to 0cm
 {\hss\
   {\includegraphics{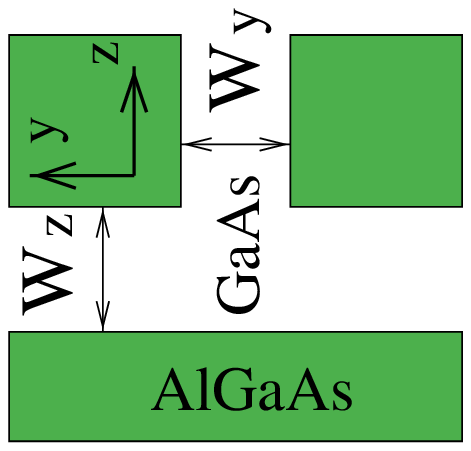}
   }
  \hss}

 }
\caption{Theoretically calculated (from Eqs. (2)-(4)) 1D
Coulomb interaction in a T-shaped QWR system in momentum space.
Results of different wire widths are calculated and
shown together. In the inset
we show the T-shaped intersection of two quantum wells in cross section.
}
\end{figure}
%-------------------------------
%--------------------------------
\begin{figure}
 \vbox to 6cm {\vss\hbox to 5.5cm
 {\hss\
   {\includegraphics{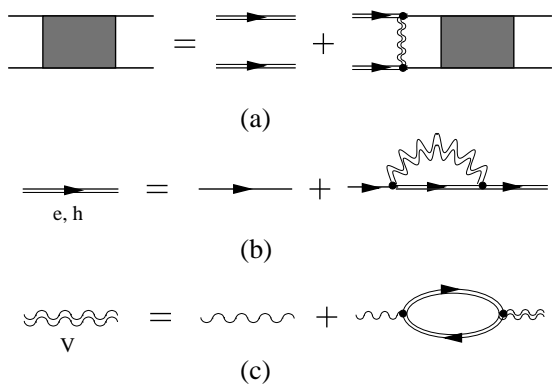}
   }
  \hss}
 }
\caption{
Many-body Feynman diagrams used in the paper with the single (double)
solid line representing the bare (dressed) electron or hole
Green's function and
the single (double) wavy line representing the bare (dressed) Coulomb
interaction: (a) the excitonic Bethe-Salpeter equation; (b) the single-loop
self-energy (in the so-called GW approximation) defining the dressed Green's
function; (c) the RPA dressing of the Coulomb interaction (treated in the
plasmon-pole approximation in our calculation).
}
\label{bse}
\end{figure}
%-------------------------------
%-------------------------------------
\begin{figure}

 \vbox to 6.3cm {\vss\hbox to 6cm
 {\hss\
   {\includegraphics{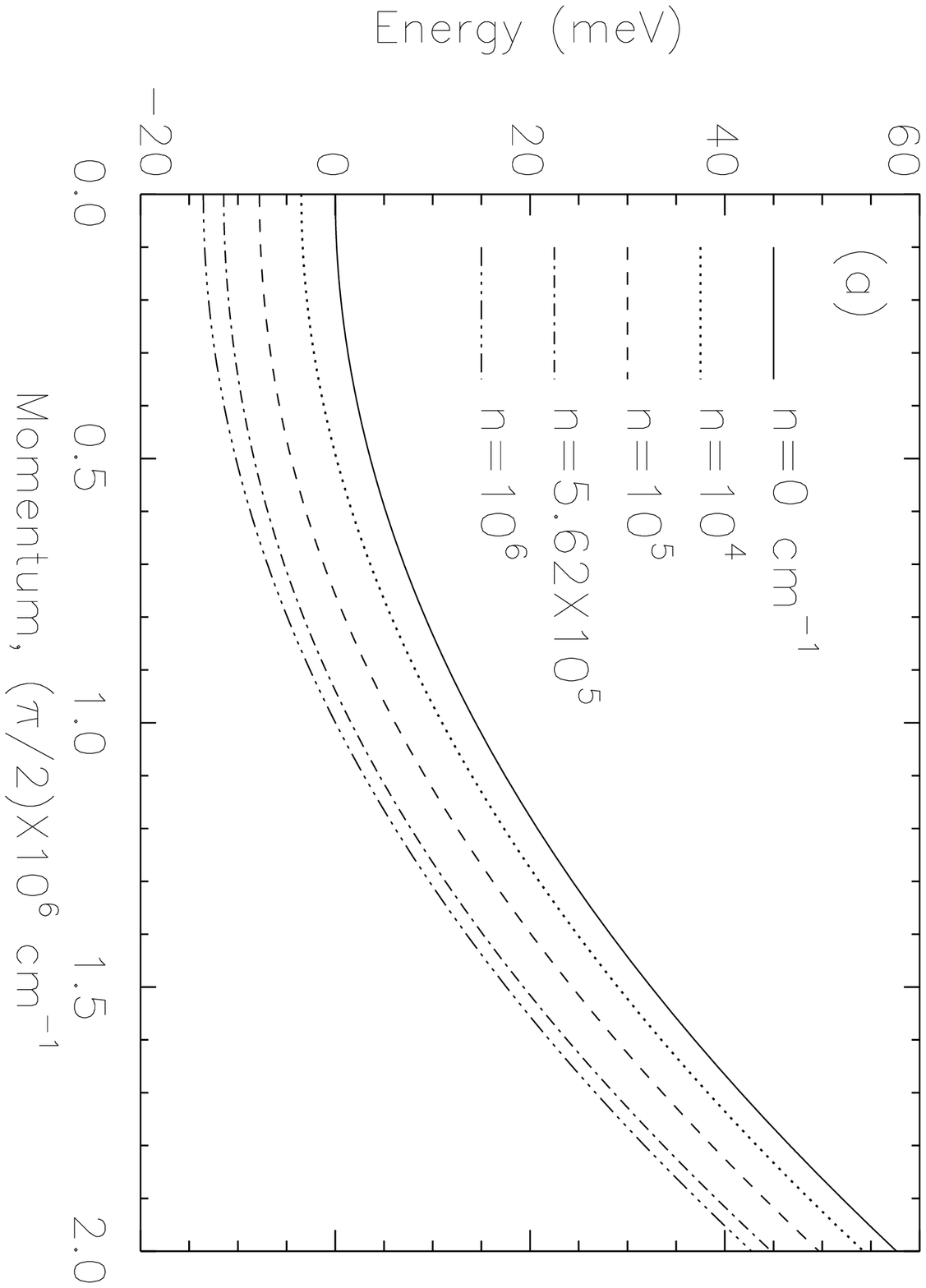}
   }
  \hss}
 }

 \vbox to 6cm {\vss\hbox to 6cm
 {\hss\
   {\includegraphics{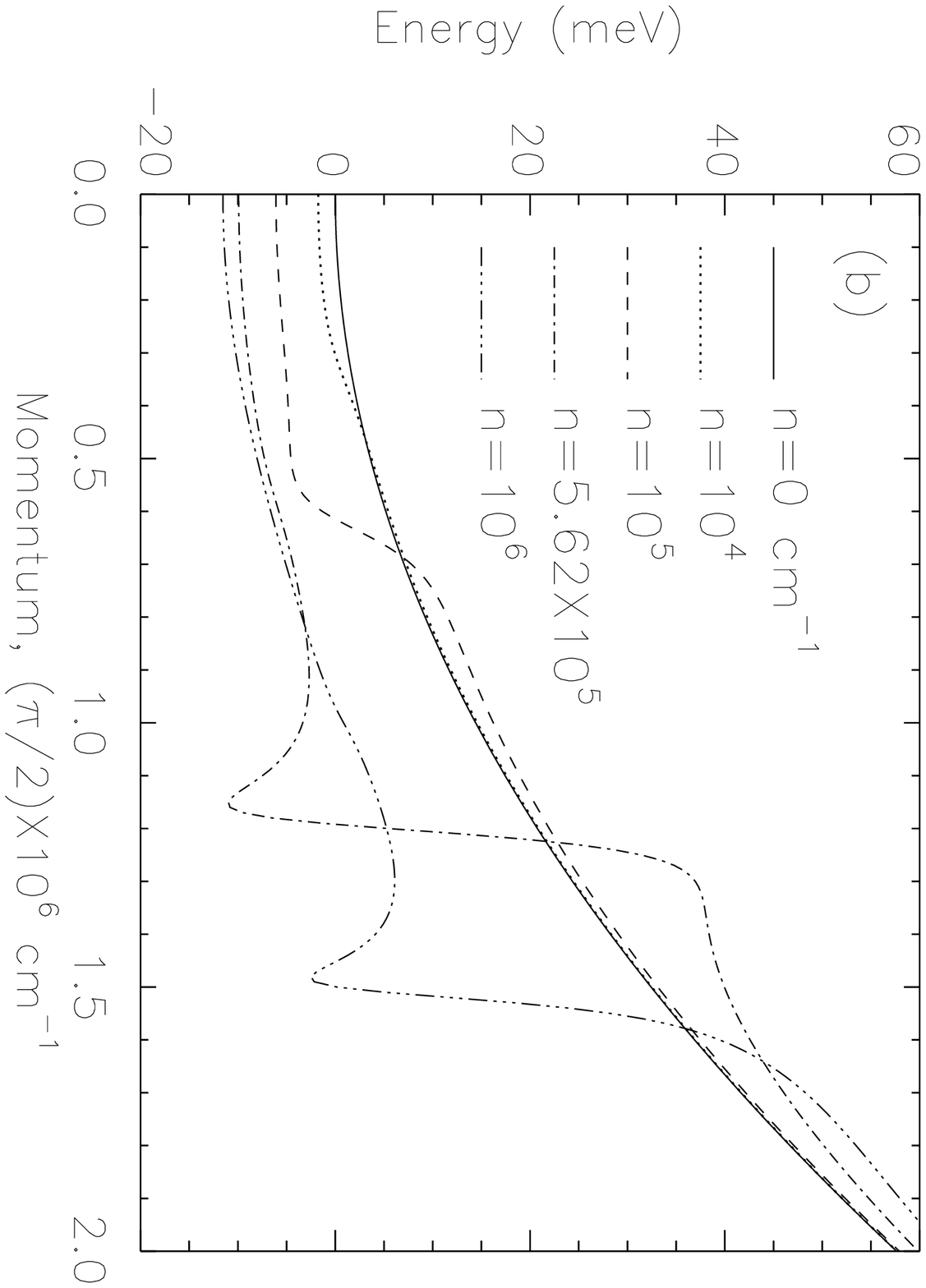}
   }
  \hss}
 }

 \vbox to 6cm {\vss\hbox to 6cm
 {\hss\
   {\includegraphics{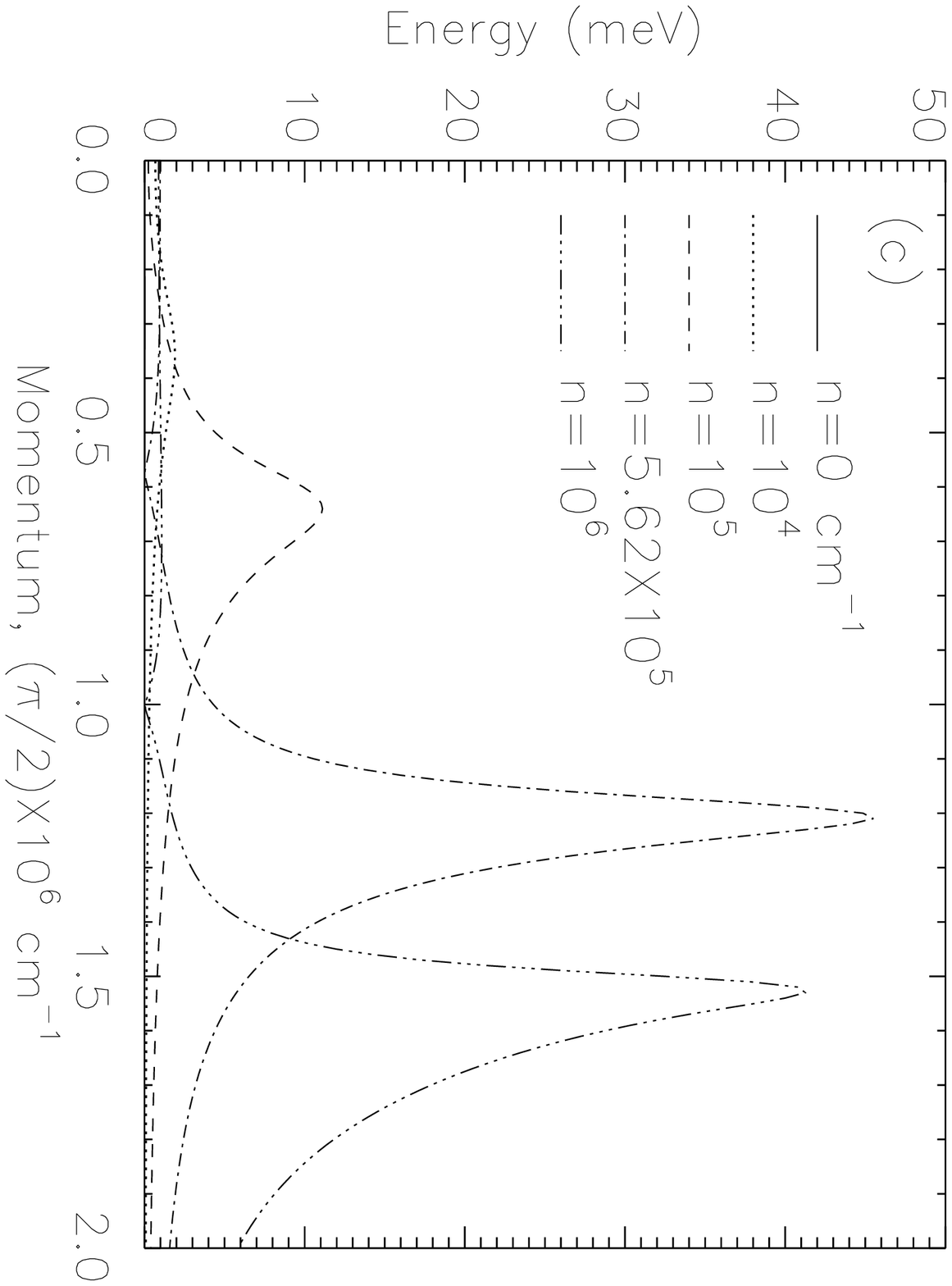}
   }
  \hss}
 }

\caption{
(a) Conduction band energy, $\xi_{e,k}-E_g^0$,
calculated in the GW approximation with screened interaction
approximated by the quasi-static-PPA.
(b) and (c) are respectively the real and imaginary parts
of band energy calculated in the dynamically screened GW approximation
for the same system as (a).
The calculation is carried out in the symmetric T-shaped QWR system 
with $W_y=W_z=7$ nm including 
finite temperature ($T=10$ K) and finite (phenomenological) impurity
scattering ($\gamma=0.5$ meV) effects.
}
\end{figure}
%-------------------------------------
%-------------------------------------
\begin{figure}

 \vbox to 6.3cm {\vss\hbox to 6cm
 {\hss\
   {\includegraphics{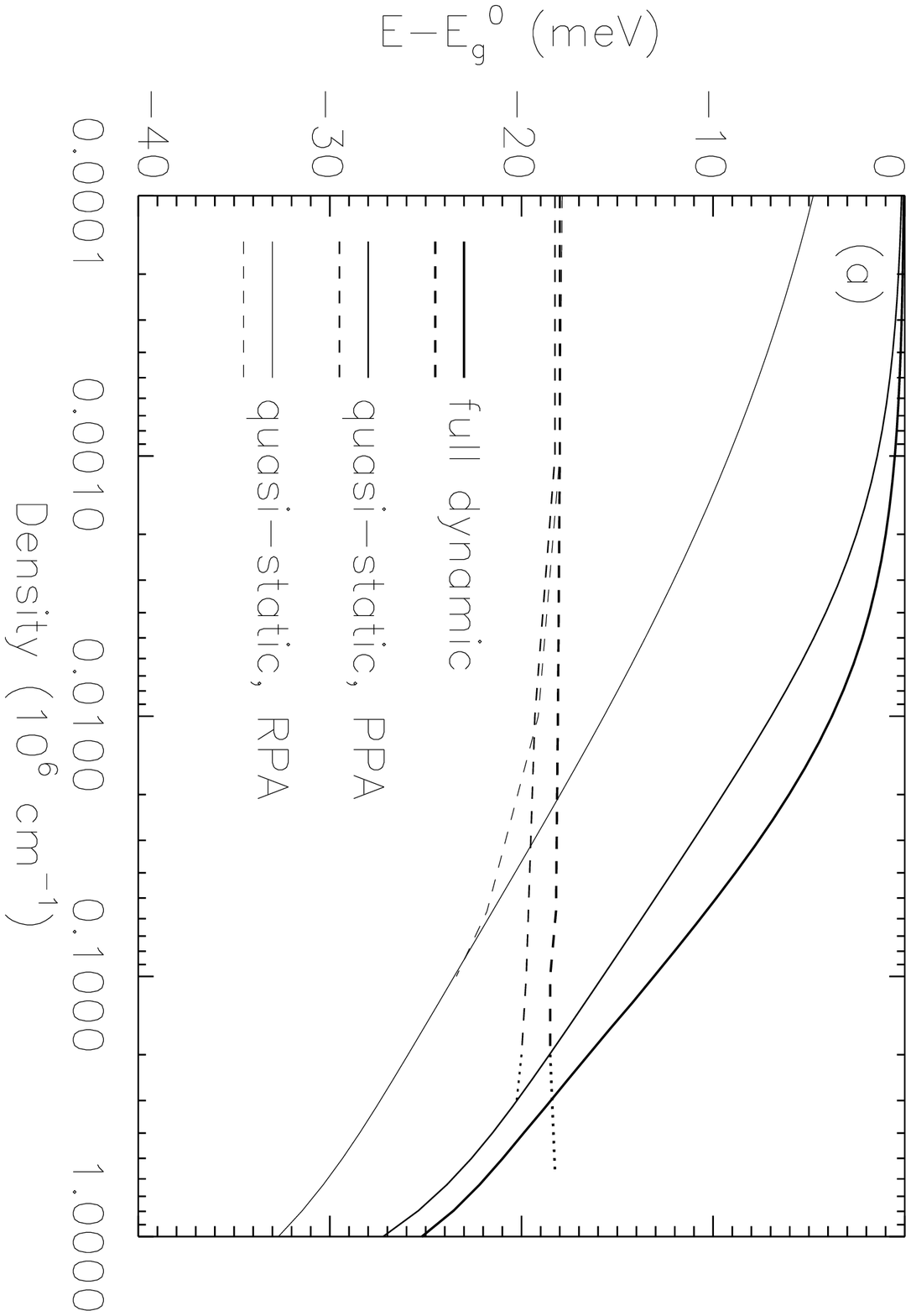}
   }
  \hss}
 }

 \vbox to 6.3cm {\vss\hbox to 6cm
 {\hss\
   {\includegraphics{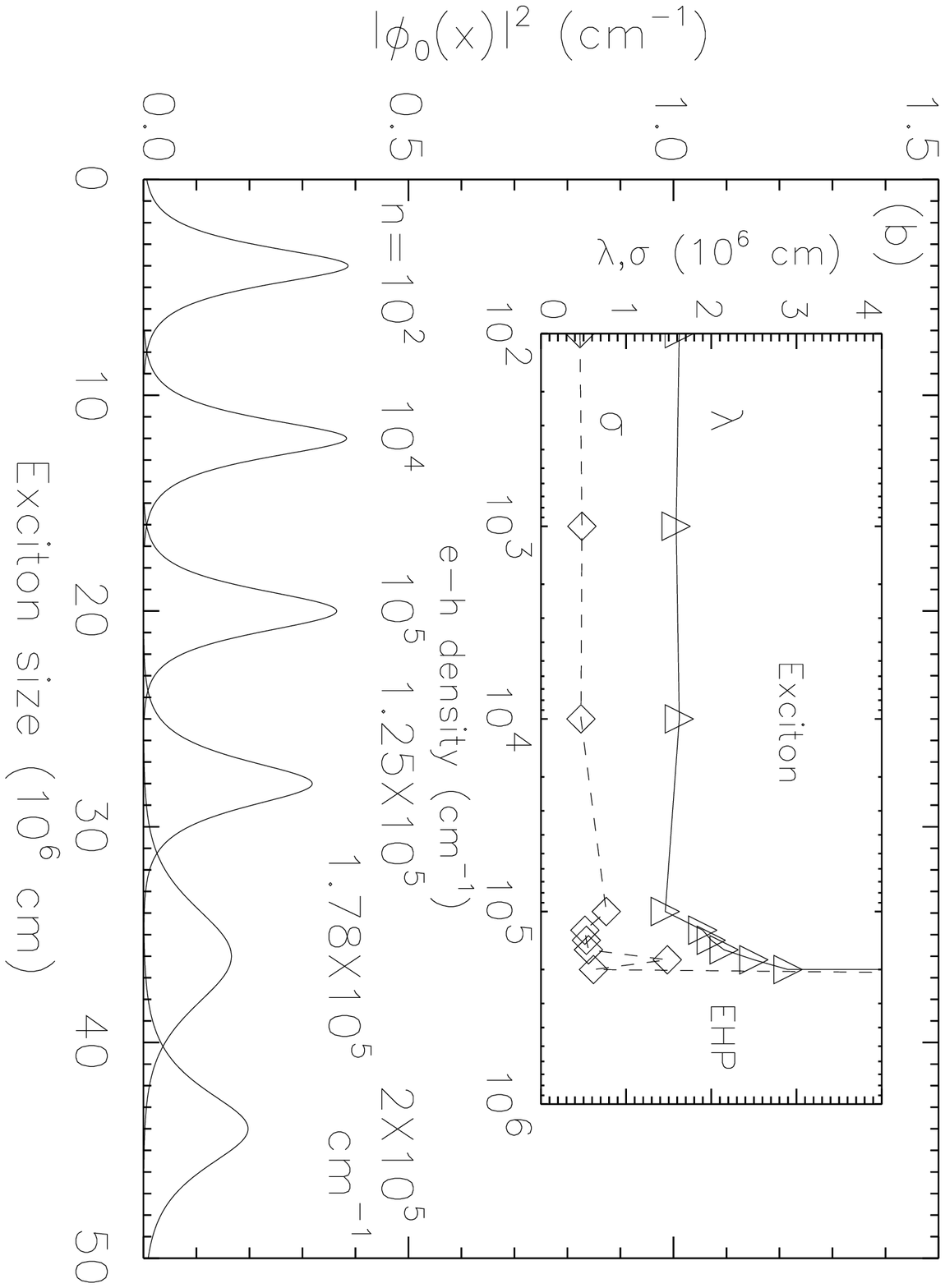}
   }
  \hss}
 }

\caption{(a)
Separately variationally calculated exciton energy (dashed lines)
and BGR of the EHP (solid lines)
as a function of photoexcitation density, in three different
approximations as indicated in the plot by different line widths.
Note that when the density is larger than $2\times 10^5$ cm$^{-1}$, 
the variational
method (introduced in Sec. II-E) fails to give a good exciton
energy (see the text) and the dotted lines are the exciton peak positions
of the corresponding absorption spectra by solving the BSE (Fig. 5).
(b) The 1$s$ exciton ground state
wavefunction obtained in the variational method through effective Hamiltonian
(Eqs. (35) and (36)) in the dynamical (PPA) screening calculation 
for various electron-hole densities.
Inset: the variational parameters, $\lambda$ and $\sigma$, for the 1$s$ 
exciton ground state trial
wavefunction with respect to the photoexcitation density in 
logarithm scale. When the 
density is near the Mott density ($n_c\sim 3\times 10^5$ cm$^{-1}$),
both $\lambda$ and $\sigma$ increase sharply
and the wave function becomes totally broadened.
}
\end{figure}
%-------------------------------------
%-------------------------------
\begin{figure}

 \vbox to 6cm {\vss\hbox to 6cm
 {\hss\
   {\includegraphics{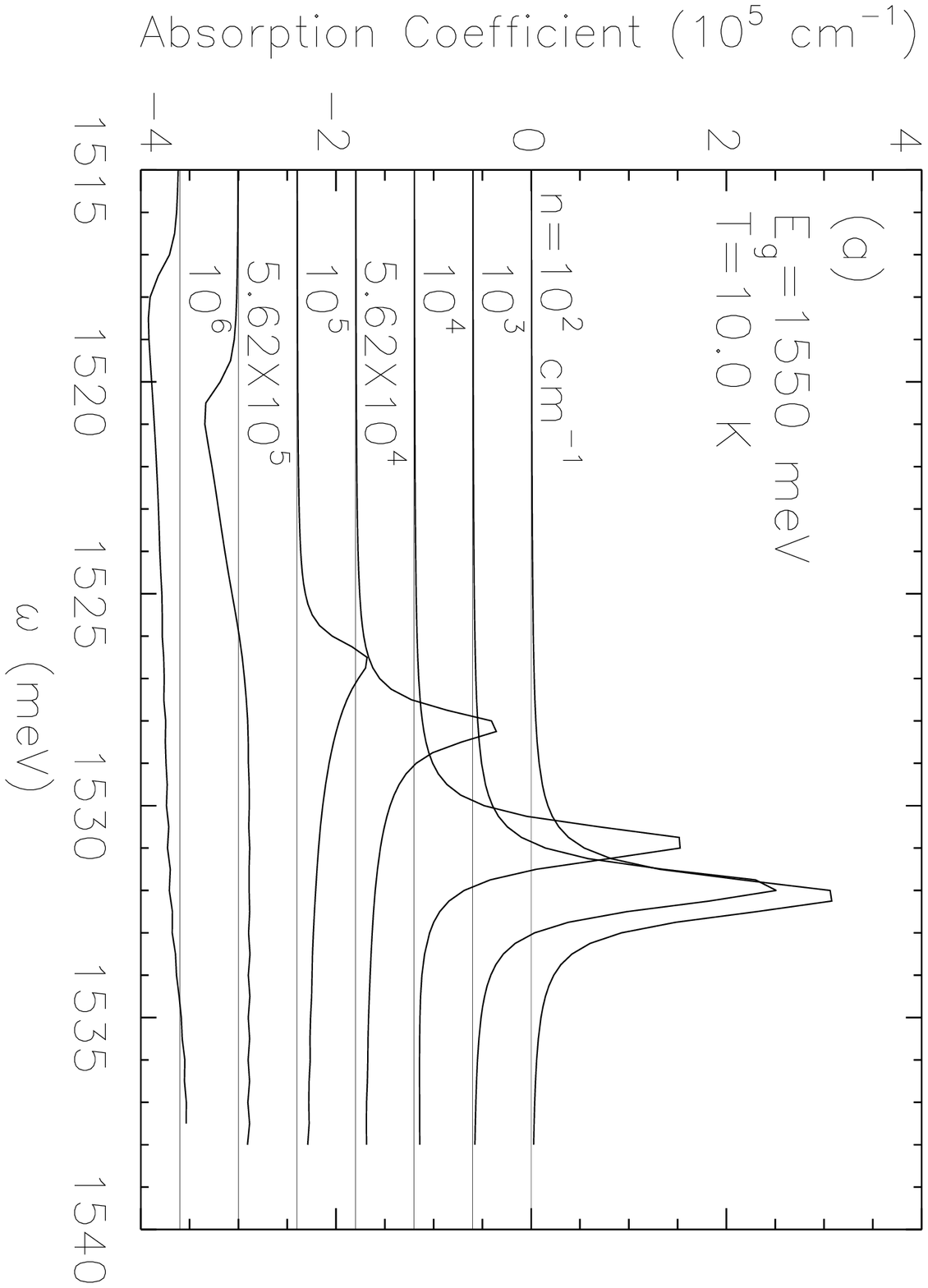}
   }
  \hss}
 }
 \vbox to 6cm {\vss\hbox to 6cm
 {\hss\
   {\includegraphics{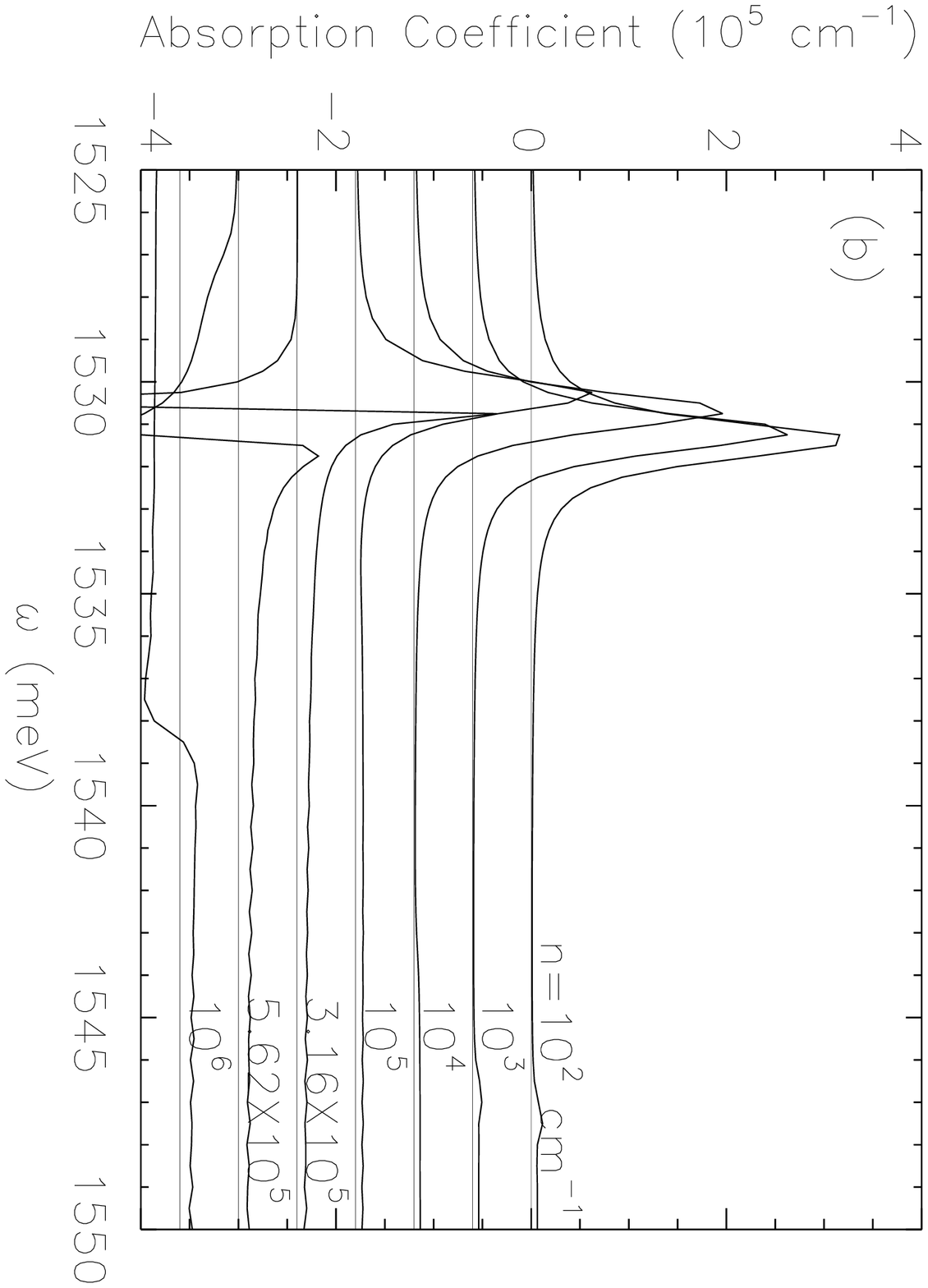}
   }
  \hss}
 }
 \vbox to 6cm {\vss\hbox to 6cm
 {\hss\
   {\includegraphics{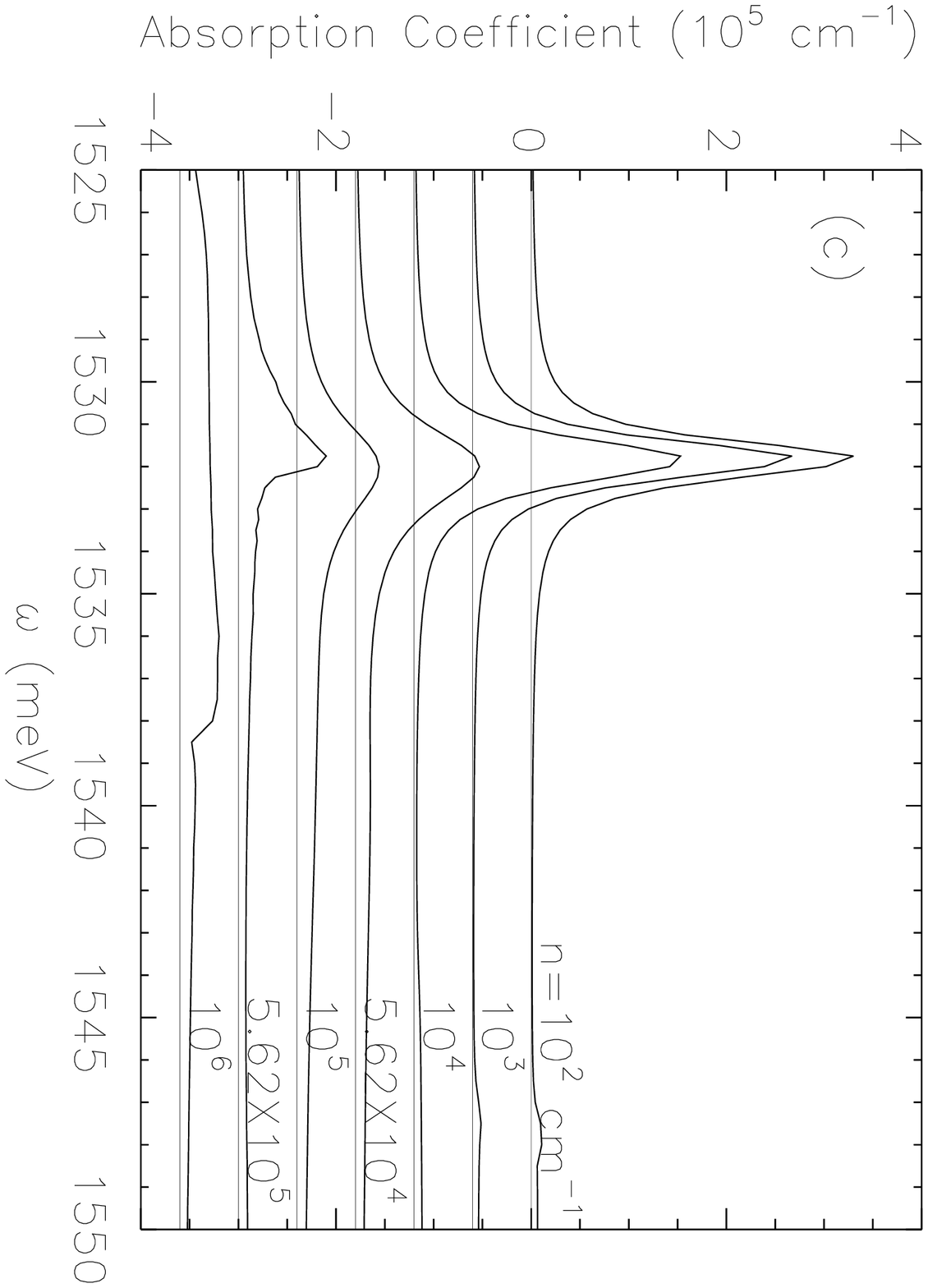}
   }
  \hss}
 }
 \vbox to 6cm {\vss\hbox to 6cm
 {\hss\
   {\includegraphics{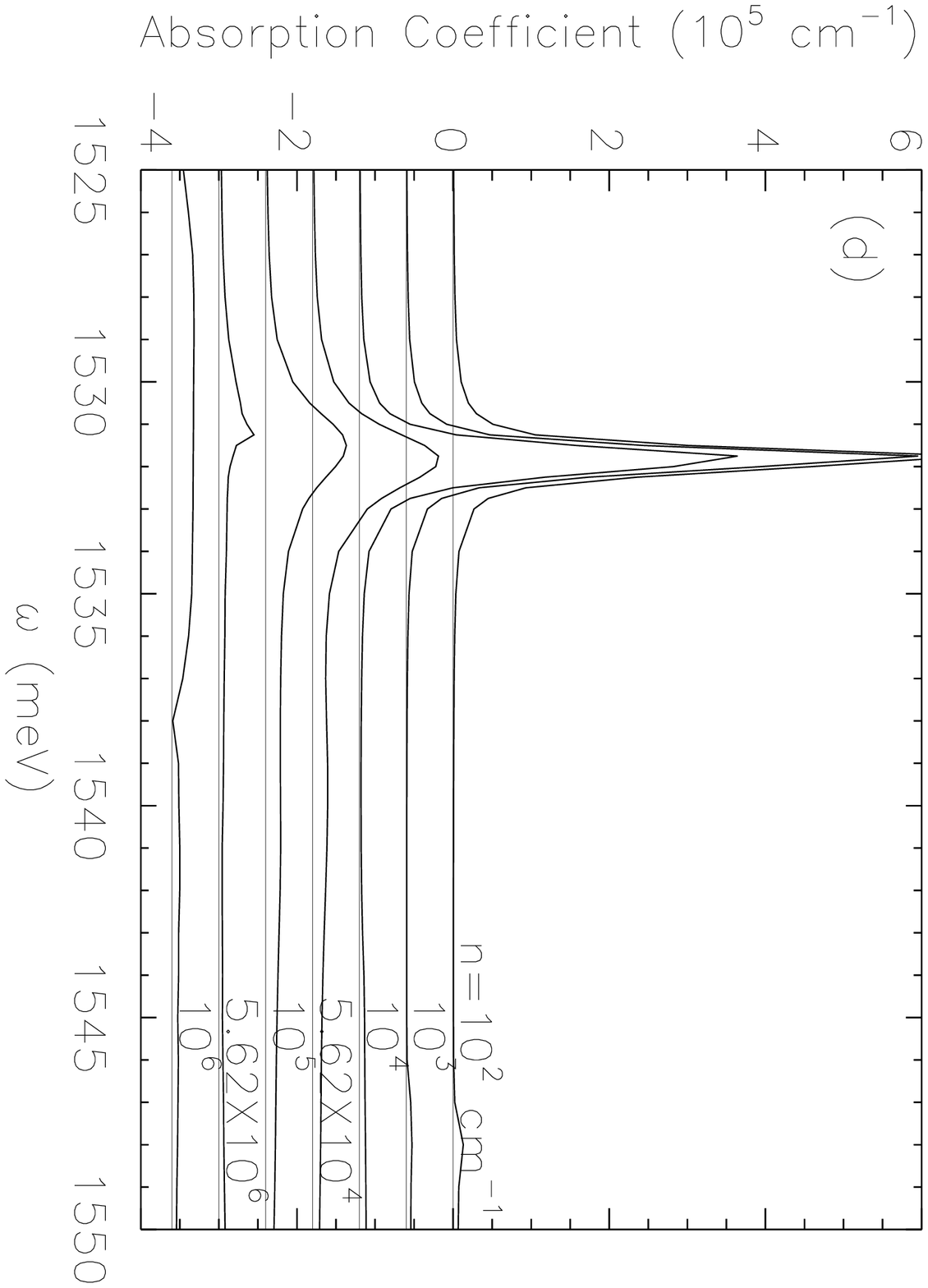}
   }
  \hss}
 }

\caption{
Calculated absorption and gain spectra for various photoexcitation densities
by solving the Bethe-Salpeter equation in three different approximations
for screening:
(a) the quasi-static-RPA, (b) the quasi-static-PPA, (c) and (d) 
the full dynamical
(PPA) calculation. The system parameters of (a)-(d) 
are the same as used in Fig. 3, except for the smaller $\gamma$ 
(=0.2 meV) used in (d).
}
\end{figure}
%----------------------------------
%-------------------------------
\begin{figure}

 \vbox to 6.3cm {\vss\hbox to 6cm
 {\hss\
   {\includegraphics{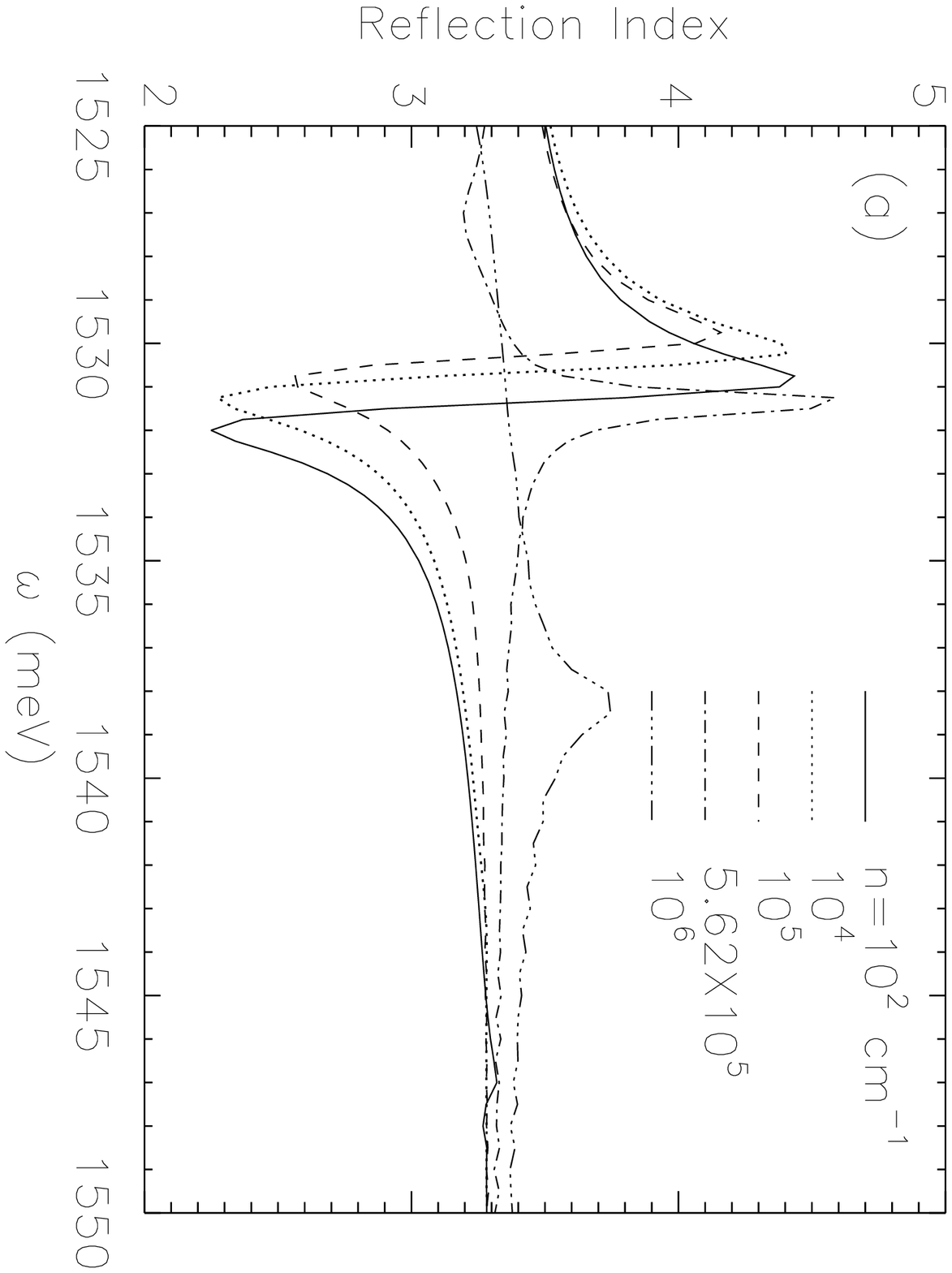}
   }
  \hss}
 }
 \vbox to 6.3cm {\vss\hbox to 6cm
 {\hss\
   {\includegraphics{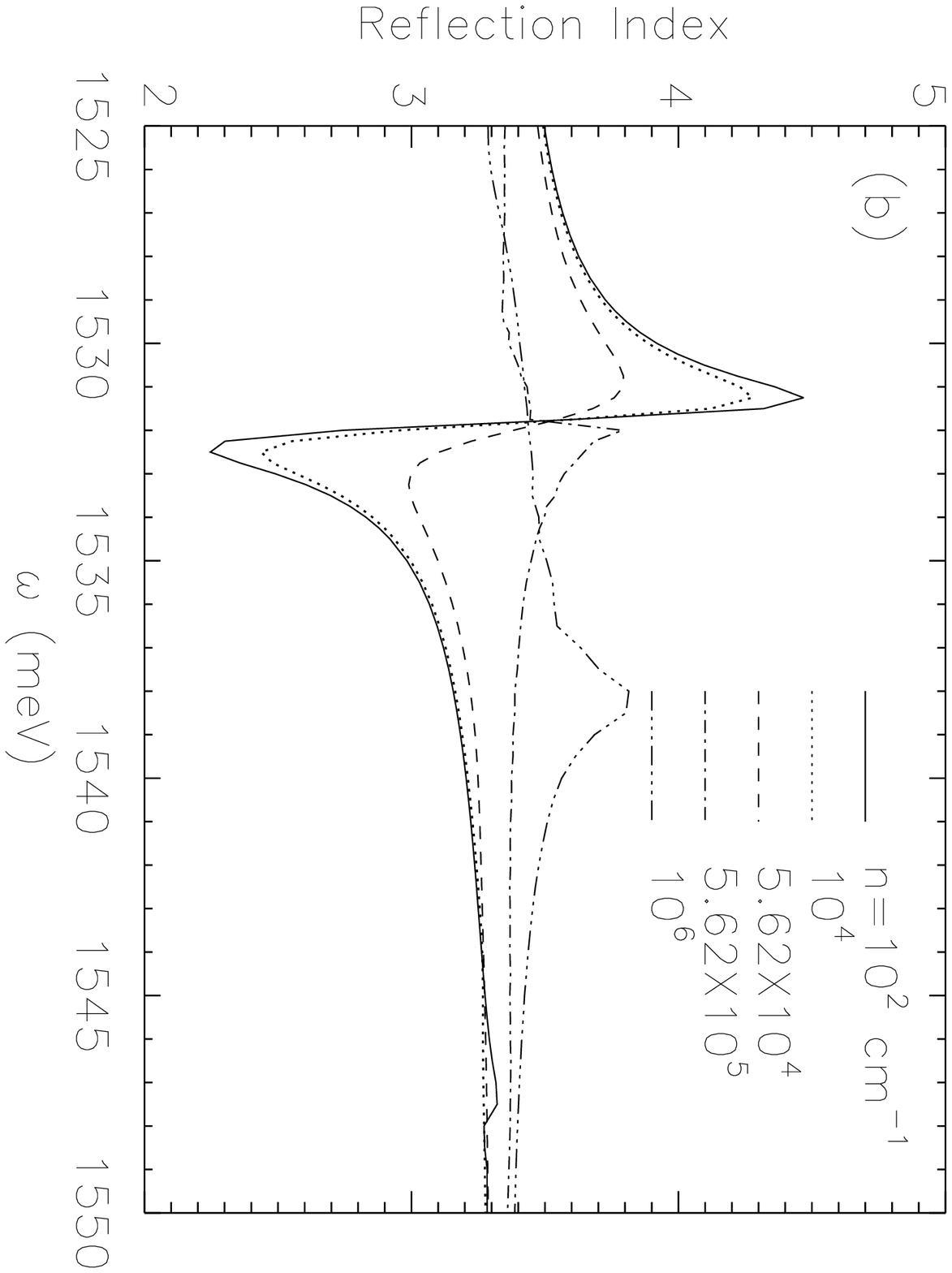}
   }
  \hss}
 }
\caption{
Calculated refractive index for various photoexcitation densities in
both (a) the quasi-static-PPA and (b) the dynamical (PPA) approximation
for interaction screening. 
}
\end{figure}
%----------------------------------
%----------------------------------
\begin{figure}
 \vbox to 6.5cm {\vss\hbox to 6cm
 {\hss\
   {\includegraphics{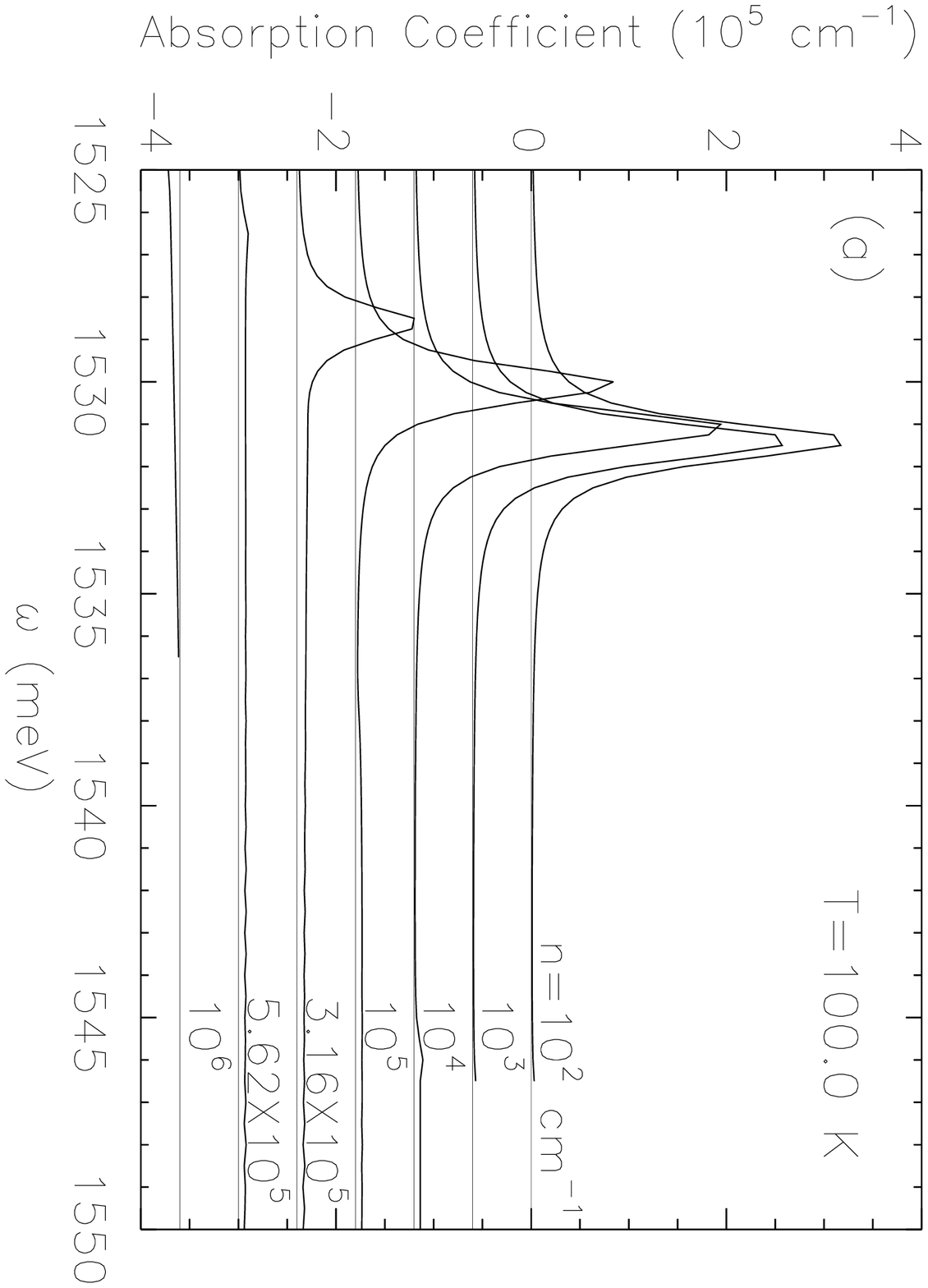}
   }
  \hss}
 }
 \vbox to 6.5cm {\vss\hbox to 6cm
 {\hss\
   {\includegraphics{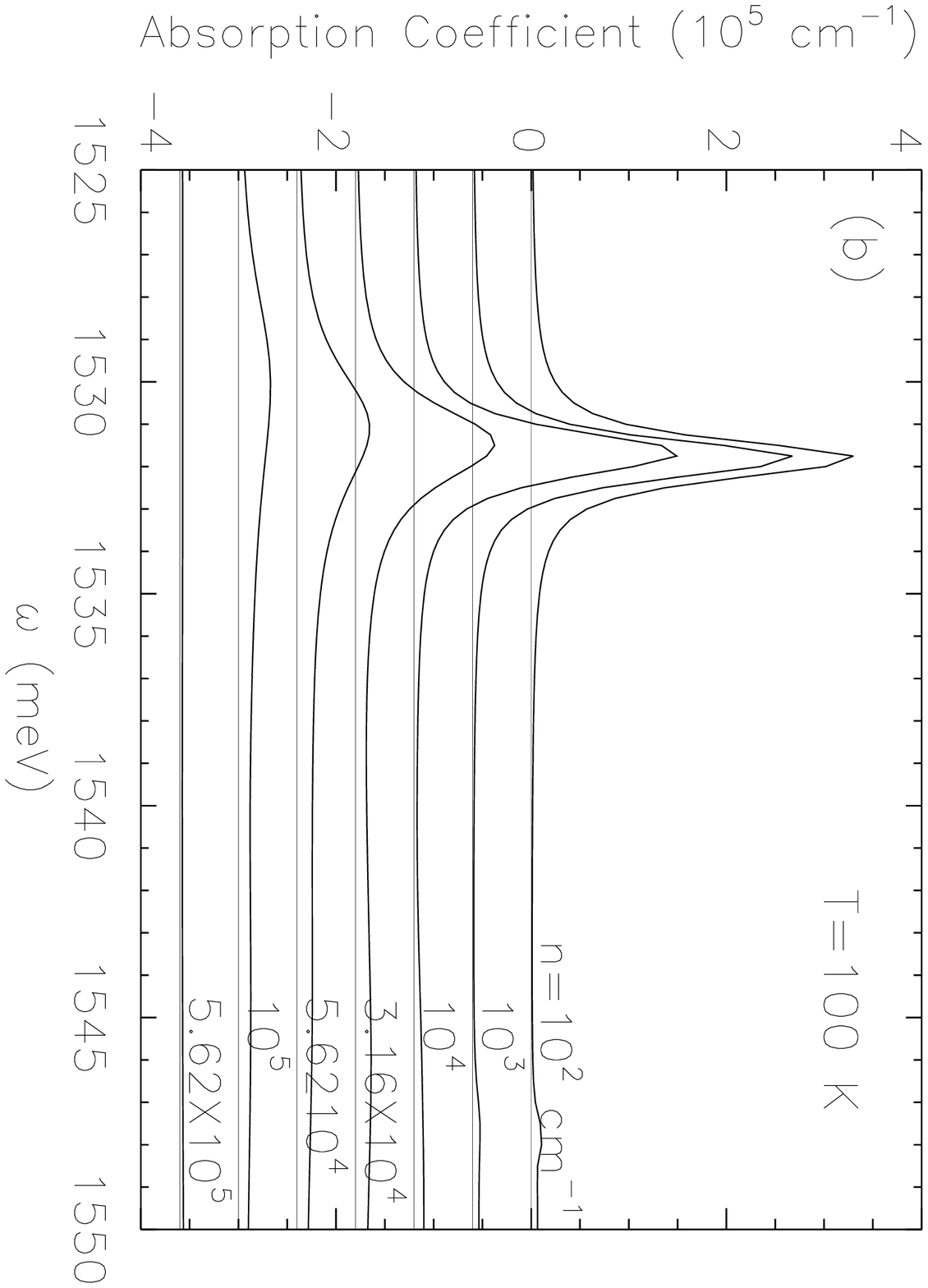}
   }
  \hss}
 }
\caption{
Absorption and gain spectra obtained by solving the Bethe-Salpeter
equation in (a) the quasi-static-PPA
and (b) the full dynamical (PPA) approximation for screening at high
temperature ($T=100$ K). Other system parameters are the same as used 
in Fig. 3.
}
\end{figure}
%----------------------------------
\end{document}